\documentclass[aps,pra,reprint,superscriptaddress,fleqn]{revtex4-2}
\usepackage[utf8]{inputenc}

\usepackage{amsmath}
\usepackage{amssymb}
\usepackage{amsfonts}
\usepackage{graphicx}
\usepackage{svg}
\usepackage[unicode=true,
 bookmarks=true,bookmarksnumbered=false,bookmarksopen=false,
 breaklinks=false,pdfborder={0 0 1},backref=false,colorlinks=true]
 {hyperref}
\hypersetup{
 linkcolor=magenta, urlcolor=blue, citecolor=blue, pdfstartview={FitH}, unicode=true}
\usepackage{xcolor}

\begin{document}
\title{Collective dynamics and long-range order in thermal neuristor networks}

\author{Yuan-Hang Zhang}
\email{email: yuz092@ucsd.edu}
\affiliation{Department of Physics, University of California San Diego, La Jolla, CA 92093}

\author{Chesson Sipling}
\affiliation{Department of Physics, University of California San Diego, La Jolla, CA 92093}

\author{Erbin Qiu}
\affiliation{Department of Physics, University of California San Diego, La Jolla, CA 92093}
\affiliation{Department of Electrical and Computer Engineering, University of California San Diego, La Jolla, CA 92093}

\author{Ivan K. Schuller}
\affiliation{Department of Physics, University of California San Diego, La Jolla, CA 92093}

\author{Massimiliano Di Ventra}
\email{email: diventra@physics.ucsd.edu}
\affiliation{Department of Physics, University of California San Diego, La Jolla, CA 92093}

\begin{abstract}
In the pursuit of scalable and energy-efficient neuromorphic devices, recent research has unveiled a novel category of spiking oscillators, termed ``thermal neuristors''. These devices function via thermal interactions among neighboring vanadium dioxide resistive memories, emulating biological neuronal behavior. Here, we show that the collective dynamical behavior of networks of these neurons showcases a rich phase structure, tunable by adjusting the thermal coupling and input voltage. Notably, we identify phases exhibiting long-range order that, however, does not arise from criticality, but rather from the time non-local response of the system. 
In addition, we show that these thermal neuristor arrays achieve high accuracy in image recognition and time series prediction through reservoir computing, without leveraging long-range order. Our findings highlight a crucial aspect of neuromorphic computing with possible implications on the functioning of the brain: 
criticality may not be necessary for the efficient performance of neuromorphic systems in certain computational tasks.

\end{abstract}
\maketitle

\section{Introduction}

Neuromorphic computing, a field inspired by brain functionality, represents a powerful approach to tackle a wide range of information processing tasks that are not instruction-based, such as those typical of artificial intelligence and machine learning   \cite{mead1990neuromorphic, markovic2020physics, schuman2022opportunities}. Unlike traditional computers that use the von Neumann architecture, separating memory and computing, neuromorphic systems utilize artificial neurons and synapses. These components can be implemented using diverse physical systems, such as photonics \cite{shastri2021photonics}, spintronics \cite{grollier2020neuromorphic}, resistive switching materials \cite{del2018challenges, qiu2024reconfigurable}, and electrochemical devices \cite{van2017non}.

In neuromorphic systems, regardless of the underlying physical framework, information processing is executed via a spiking neural network \cite{maass1997networks}. Neurons in this network emit spikes in response to specific external stimuli. These spikes travel through synapses, either exciting or inhibiting downstream neurons. During training for a particular task, synaptic weights are iteratively updated, guided by either biologically-inspired algorithms like spike timing-dependent plasticity \cite{caporale2008spike} and evolutionary algorithms \cite{pavlidis2005spiking} or adaptations of traditional machine learning algorithms like backpropagation \cite{tavanaei2019deep}. 

The collective, as opposed to the individual behavior of the neurons in the network, facilitates the aforementioned tasks. This collective behavior may also be essential for the functioning of the animal brain. For instance, the ``critical brain hypothesis'' suggests that the brain operates in a state of ``criticality''; namely, it is poised at a transition point between different phases \cite{chialvo2010emergent, hesse2014self, beggs2012being, di2018landau, o2022critical}. This critical state is believed to be optimal for the brain's response to both internal and external stimuli, due to its structural and functional design. Yet, despite the popularity of the hypothesis, questions and doubts remain, and some argue that the brain is not truly critical or not critical at all \cite{beggs2012being, priesemann2018can, wilting201925, beggs2022addressing}. 

In our present study, we do not aim to directly tackle the critical brain hypothesis. Rather, we approach the subject from a different angle: we examine a neuromorphic system that exhibits brain-like features. With similar working principles, one can then naturally extend the critical brain hypothesis to neuromorphic systems and question whether spiking neural networks also function at a critical state. This topic remains contentious, and arguments supporting \cite{brochini2016phase, scarpetta2018hysteresis} and opposing \cite{cramer2020control} the notion have been reported, each presenting slightly different definitions and perspectives.

In this work, we show that a neuromorphic system may support long-range ordered (LRO) phases, without criticality. The origin of this LRO is the time non-local (memory) response of the system to external perturbations. On the other hand, we show that such LRO is not necessary for certain computational tasks, such as classification and time series predictions. These results may provide some hints   
on the functioning of biological brains. 
\begin{figure*}
    \centering
    \includegraphics[width=1\linewidth]{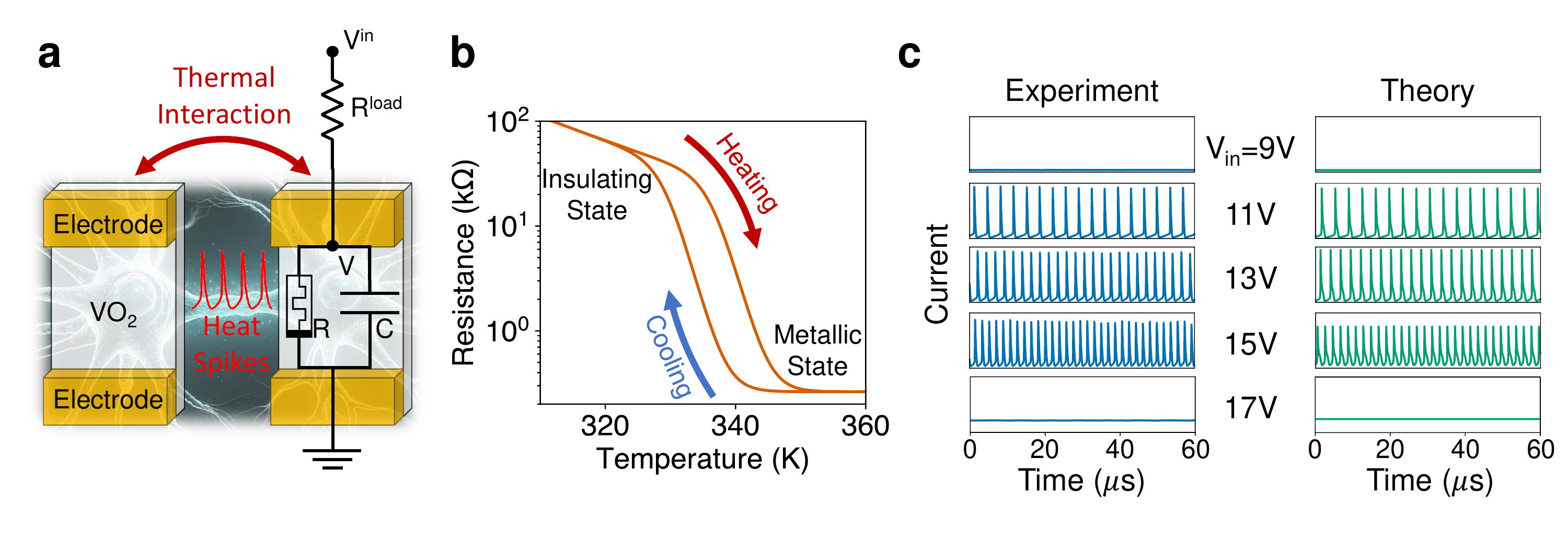}
    \caption{(a) Schematic and circuit diagram of two neighboring thermal neuristors. Each neuristor is modeled as an RC circuit, which undergoes stable spiking oscillations with proper external input. Neighboring neuristors are electrically isolated but communicate with each other through thermal interactions. (b) The resistance-temperature characteristic of the VO$_2$ film, denoted by the variable resistor R in panel (a). VO$_2$ exhibits an insulator-to-metal transition at approximately 340 K, characterized by distinct heating and cooling trajectories, thus forming a hysteresis loop.  (c) Illustration of stable spiking oscillations in a single neuristor across various input voltages, with the y-axis range for each plot set between 0 and 5 mA. Numerical simulations based on Eqs.~(\ref{eq:current}) and~(\ref{eq:heat}) align well with experimental data, demonstrating stable spiking patterns within a certain input voltage range and an increase in spiking frequency proportional to the input voltage.}
    \label{fig:schematic}
\end{figure*}

As a specific example, we consider a neuromorphic system comprised of thermal neuristors \cite{qiu2023stochastic, qiu2024reconfigurable}, based on vanadium dioxide (VO$_2$) spiking oscillators that communicate via heat signals. The properties of the individual oscillators (which take advantage of the hysteric metal-insulator transition of VO$_2$) and their mutual interactions have been experimentally validated earlier \cite{qiu2023stochastic, qiu2024reconfigurable}. These earlier studies form the basis of our numerical model of a large-scale network, which allows us to numerically analyze the collective dynamics of the system. We find that the different phases can be tuned by varying the thermal coupling between the neurons and the input voltage. We apply this system to image recognition tasks using reservoir computing \cite{tanaka2019recent} and explore the relationship between performance and collective dynamics. We find 
that LRO does not necessarily enhance the performance in tasks like image recognition, a result in line with the findings of Ref.  \cite{cramer2020control}.

\section{Results}
VO$_2$-based oscillators have been utilized as artificial neurons in many previous studies \cite{yi2018biological, belyaev2019spiking, velichko2019model, velichko2020reservoir, yuan2022calibratable, qiu2023stochastic, qiu2024reconfigurable}, each featuring slightly different designs, mechanisms, and applications. In particular, we focus on thermal neuristors, a concept pioneered in \cite{qiu2024reconfigurable}, which effectively reproduces the behavior of biological neurons. These neuristors are not only straightforward to manufacture experimentally but also exhibit advantageous properties such as rapid response times and low energy consumption.

Fig.~\ref{fig:schematic}(a) presents the design and circuitry of the thermal neuristor, featuring a thin VO$_2$ film connected in series to a variable load resistor. VO$_2$ undergoes an insulator-to-metal transition (IMT) at approximately 340 K \cite{zylbersztejn1975metal}, with different resistance-temperature heating and cooling paths, which leads to a hysteresis loop, as depicted in Fig.~\ref{fig:schematic}(b). Additionally, the system includes a parasitic capacitance resulting from the cable connections,
which is vital for the neuristor's operation.

The behavior of the circuit displayed in Fig.~\ref{fig:schematic}(a) closely resembles a leaky integrate-and-fire neuron \cite{gerstner2002spiking}. The capacitor $C$ is charged up by the voltage source, $V^\mathrm{in}$, and slowly leaks current through $R$. When the voltage across VO$_2$ reaches a threshold, joule heating initiates the IMT, drastically reducing resistance in the VO$_2$ which causes $C$ to discharge, leading to a current spike. At the same time, the reduced resistance leads to reduced joule heating, which is then insufficient to maintain the metallic state, causing the VO$_2$ film to revert to its insulating phase. This process repeats, producing consistent spiking oscillations.

We have experimentally fabricated and evaluated this system of VO$_2$-based thermal neuristors. The spiking behavior of a single neuristor is shown in Fig.~\ref{fig:schematic}(c). 
With insufficient heating, the neuristor does not switch from the insulating state whereas excessive heating keeps it perpetually in the metallic state. As a consequence, no spiking patterns emerge when the input voltage is too low or too high. Numerical simulations, using the model described in the next section, corroborate this behavior, mirroring the experimental findings.

Distinct from biological neurons that communicate via electrical or chemical signals, thermal neuristors interact through heat. As illustrated in Fig.~\ref{fig:schematic}(a), adjacent neuristors, while electrically isolated, can transfer heat via the substrate. Each current spike produces a heat spike, which spreads to nearby neuristors, reducing their IMT threshold voltage, thereby causing an excitatory interaction. Conversely, excessive heat can cause neighboring neuristors to remain metallic and cease spiking, akin to inhibitory interactions between neurons. Further experimental insights on neuristor interactions are detailed in Appendix B the supplementary information (SI). 

Although we have experimentally shown that a small group of thermal neuristors can mirror the properties of biological neurons, effective computations require a vast network of interacting neurons. Before building a complex system with many neuristors, we first simulate a large array of thermal neuristors, providing a blueprint for future designs.

\subsection{Theoretical model}
\label{sec:theory}

\begin{figure*}
    \centering
    \includegraphics[width=1\linewidth]{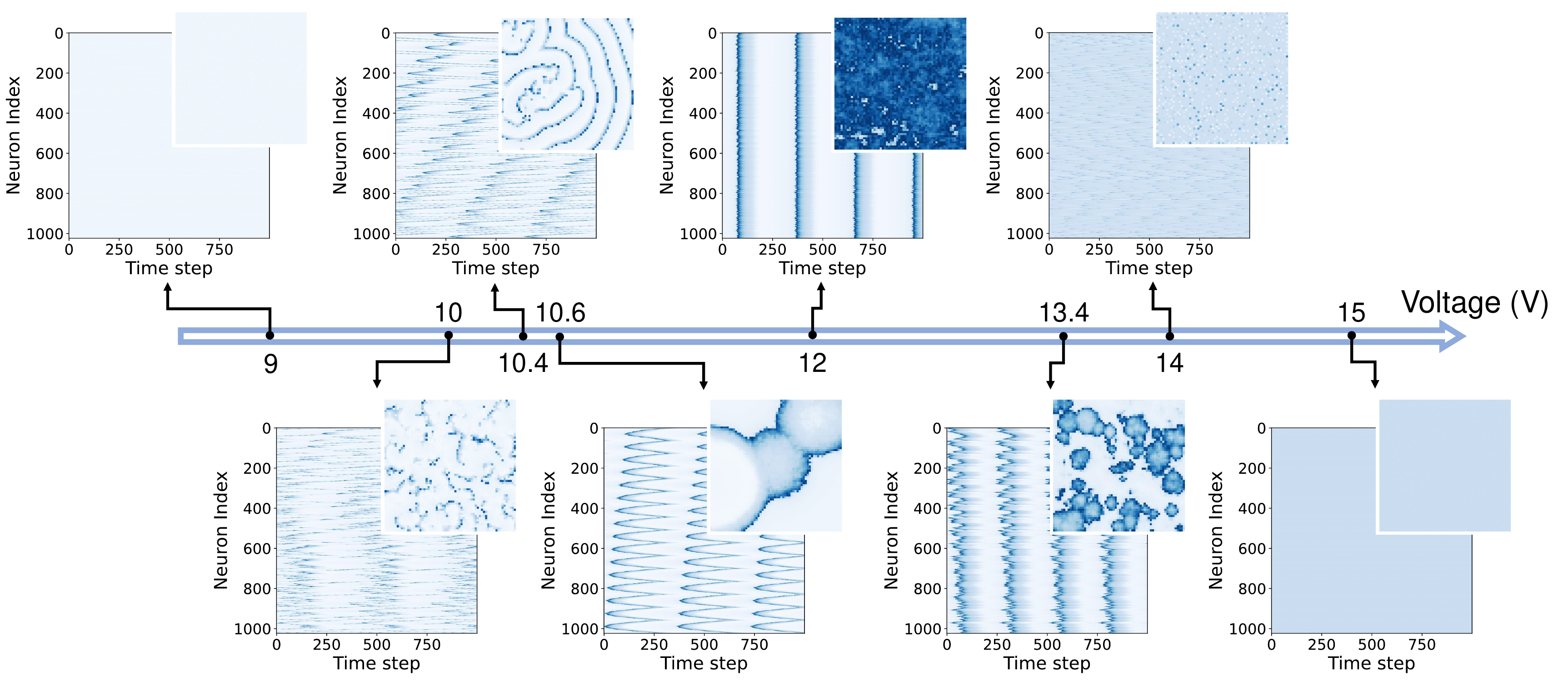}
    \caption{Snapshots of different oscillation patterns in a $64\times 64$ array of thermal neuristors. In each panel, color indicates current level: white signifies no current, while shades of blue denote current spikes. The main panels show collective current-time plots for the first 1024 neuristors (concatenated from the first 16 rows), and each inset captures a specific moment in the $64\times 64$ array. The system exhibits no activity at very low input voltages. As the voltage increases, a sequence of dynamic phases unfolds, including correlated clusters (10 V and 13.4 V), system-wide waves (10.4 V and 10.6 V), synchronized rigid states (12 V), and uncorrelated spikes (14 V), culminating again in inactivity at excessively high voltages. The thermal capacitance, $C_\mathrm{th}$, is fixed at the experimentally estimated value. Detailed simulation parameters can be found in the methods section, and dynamic visualizations of these spiking patterns are available in the Supplementary movie.}
    \label{fig:snapshots}
\end{figure*}

The theoretical model builds upon the framework established in \cite{qiu2024reconfigurable}, with some minor adjustments. The system is built of identical neuristors, uniformly spaced in a regular 2-dimensional array. Their behavior is governed by the following equations:
\begin{align}
    C\frac{dV_i}{dt}=& \frac{V_{i}^{\mathrm{in}}}{R^{\mathrm{load}}_{i}}-V_i\left(\frac{1}{R_i}+\frac{1}{R^{\mathrm{load}}_{i}}\right)\label{eq:current},\\ 
    C_{\mathrm{th}}\frac{dT_i}{dt}=&\frac{V_i^2}{R_i}-S_{\mathrm{e}}(T_i-T_\mathrm{0})+S_{\mathrm{c}}\nabla^2T_i + \sigma\eta_i(t)\label{eq:heat}.
\end{align}

Equation~\eqref{eq:current} describes the current dynamics, with each variable corresponding to those shown in Fig.~\ref{fig:schematic}(a). Equation ~\eqref{eq:heat} describes the thermal dynamics, including the coupling between nearest-neighbor neuristors. Here,  $T_\mathrm{0}$ represents the ambient temperature, $C_\mathrm{th}$ is the thermal capacitance of each neuristor, $S_\mathrm{e}$ denotes the thermal conductance between each neuristor and the environment, and $S_\mathrm{c}$ refers to the thermal conductance between adjacent neuristors. $\eta_i(t)$ represents a Gaussian white noise variable for each neuristor that satisfies $\langle \eta_i(t)\eta_j(t') \rangle= \delta_{i,j}\delta(t-t')$, and $\sigma$ is the noise strength. Detailed values of these constants are provided in the methods section. $R_i$ is the resistance of the VO$_2$ film, which depends on temperature and its internal state, or memory, following the hysteresis loop depicted in Fig.~\ref{fig:schematic}(b). This memory factor is pivotal in determining the collective behavior of thermal neuristors. We utilize the hysteresis model formulated in \cite{de2002modeling}, with comprehensive details available in the methods section.

\subsection{Numerical results}

We used the theoretical model to simulate an $L\times L$ square lattice comprised of identical thermal neuristors, whose dynamics are governed by Eqs.~\eqref{eq:current} and \eqref{eq:heat}. Different input voltages $V^\mathrm{in}$ produce a diverse array of oscillation patterns, as illustrated in Fig.~\ref{fig:snapshots}. At very low (9V) or high (15V) input voltages, the system remains inactive, as found in individual neuristors. With a 12 V input voltage, synchronization develops, with nearly all neuristors spiking in unison, creating a phase of rigid states. A phase transition occurs slightly below $V^\mathrm{in}=10$ V, where clusters of correlated spikes start to form, then gradually turn into system-wide activity waves (10.4 V and 10.6 V). Another phase transition occurs slightly above $V^\mathrm{in}=13.4$ V, where the synchronized rigid oscillations start to fracture into smaller clusters until the individual spikes become uncorrelated (14 V). 

\subsection{Analytical understanding}

The emergence of a broad range of phases and long-range correlations in our system, despite only diffusive coupling existing between neurons, is a point of significant interest. Diffusive coupling is typically associated with short-range interactions, making the discovery of long-range correlations particularly intriguing.

It is well-established that long-range correlations can emerge from local interactions in various systems such as sandpiles \cite{bak1987self}, earthquake dynamics \cite{olami1992self}, forest fires \cite{drossel1992self}, and neural activities \cite{beggs2003neuronal}. These systems exhibit ``avalanches''—cascades triggered when one unit's threshold breach causes successive activations—manifesting as power-law distributions of event sizes, indicative of scale-free or near scale-free behaviors.

Such spontaneously emerging long-range correlations are often described under the framework of ``self-organized criticality'' \cite{bak1987self, olami1992self}. However, this term may be misleading. ``Criticality'' suggests a distinct boundary, characterized by a phase above and below it, as seen in the sandpile model where an appropriately defined order parameter undergoes a second-order phase transition \cite{gil1996landau, bonachela2009self}. In contrast, systems like earthquakes, while displaying power-law behaviors, do not exhibit true scale-invariance \cite{de2000self} and can be described as undergoing continuous phase transitions without clear critical boundaries \cite{bonachela2009self}.

We argue that the observed LRO in our system, similar to those in systems without genuine scale-free behaviors, is induced by memory (time non-local) effects stemming from a separation of time scales: a slow external drive contrasts sharply with fast avalanche dynamics. In our system, we identified three distinct time scales: the metallic RC time ($\tau_\mathrm{met}=R_\mathrm{met}C \sim 187$ ns), the insulating RC time ($\tau_\mathrm{ins}=R_\mathrm{ins}C\sim 7.57$ $\mu$s), and the thermal RC time ($\tau_\mathrm{th}=R_\mathrm{th}C_\mathrm{th}=C_\mathrm{th}/(S_\mathrm{c}+S_\mathrm{e}) \sim 241$ ns). We observe that $\tau_\mathrm{met}\lessapprox\tau_\mathrm{th}\ll \tau_\mathrm{ins}$. As the spiking and avalanche dynamics are primarily controlled by $\tau_\mathrm{met}$ and $\tau_\mathrm{th}$, and the driving dynamics by $\tau_\mathrm{ins}$, our system does exhibit an approximate separation of time scales. 

This separation allows us to conceptualize the slower time scale as memory, which retains long-term information about past states and remains relatively constant within the faster time scale, capable of preserving non-local temporal correlations. As a consequence, neuristors that are spatially distant are progressively coupled, resulting in long-range spatial correlations. This concept is systematically explored in a spin glass-inspired model \cite{sipling2024memory}, and similar behavior is also observed in a class of dynamical systems with memory (memcomputing machines) used to solve combinatorial optimization problems~\cite{Branching}. In Appendix A in the SI, we provide an analytical derivation of this phenomenon using a slightly simplified version of our model.

Consequently, altering the memory strength, specifically through adjustments of the thermal time scale $\tau_\mathrm{th}$ by varying $C_\mathrm{th}$ (the thermal capacitance of each neuristor), should result in changes to the oscillation patterns and the presence or absence of long-range correlations. Indeed, we find that by modifying $C_\mathrm{th}$, we can control the rate of heat dissipation, effectively influencing the memory's response time. Additionally, in Appendix C5 of the SI, we present another example where increasing the ambient temperature reduces the insulating RC time, thereby diminishing memory and minimizing long-range correlations.

\subsection{Avalanche size distribution}

\begin{figure}
    \centering
    \includegraphics[width=1\linewidth]{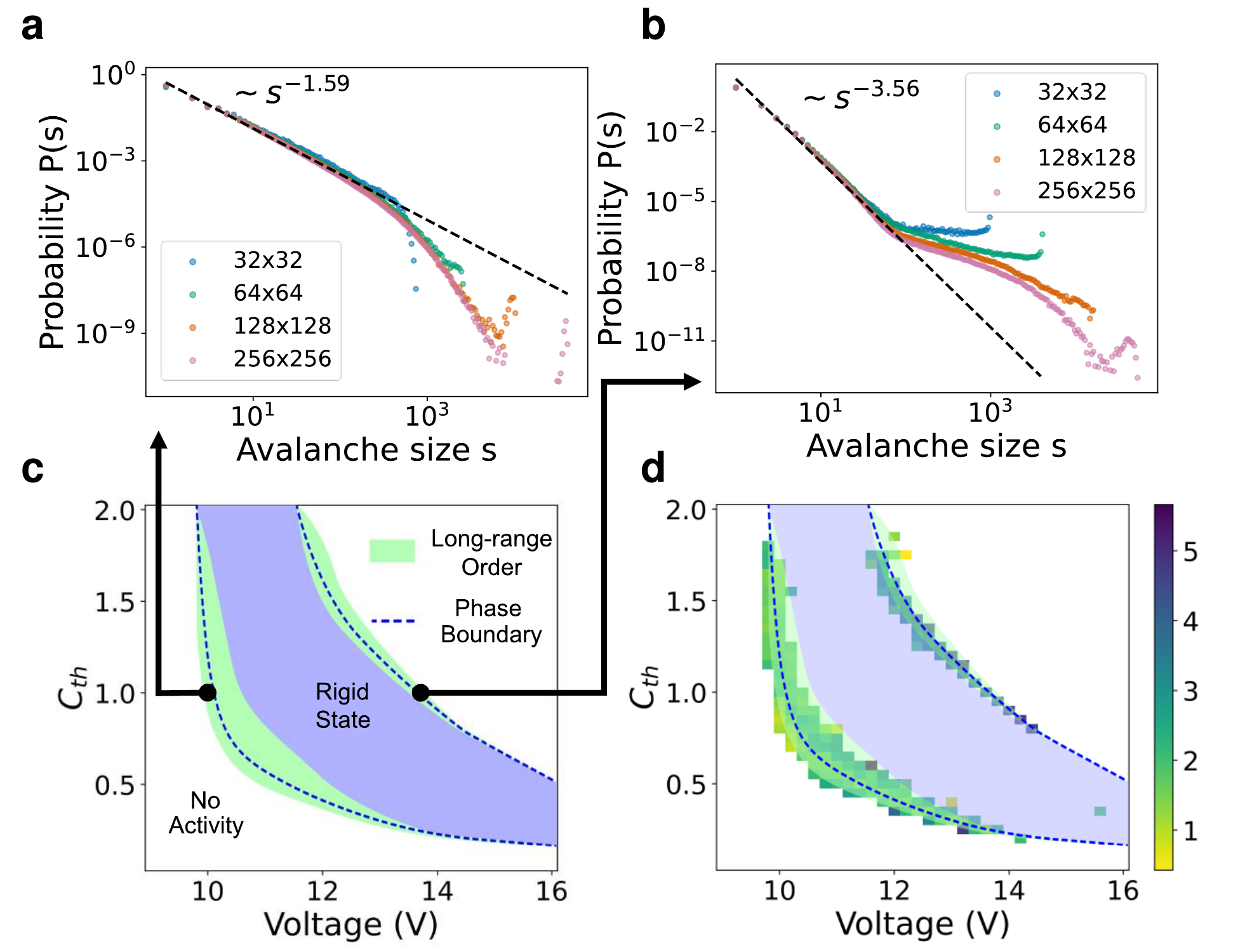}
    \caption{Avalanche size distributions and phase structures in 2D thermal neuristor arrays of different sizes. (a)(b) Two different avalanche size distributions at phase boundaries, with both distributions obtained at $C_\mathrm{th}=1$, but different input voltages ($V^\mathrm{in}=9.96$ V for (a), and 13.46 V for (b)). 
    (c) Phase diagram of the thermal neuristor array, with the y-axis depicting the relative value of C$_\mathrm{th}$ compared to its experimentally estimated level. We observe synchronized rigid states with collective spiking and quiescent states with no spikes (no activity). Near the phase boundaries, a robust power-law distribution in avalanche sizes is noted across various parameters, signaling the existence of LRO. (d) Exponents of the power-law fit of the avalanche size distributions (omitting the negative signs for clarity). The phase diagram from panel (c) is superimposed for enhanced visualization. Regions lacking a colored box signify a failed power-law fit, and exponents are capped at 6 to exclude outliers. 
    }
    \label{fig:avalanche}
\end{figure}

To verify the presence of LRO in our system, we analyzed the avalanche size distribution of current spikes. Here, we define an avalanche as a contiguous series of spiking events occurring in close spatial (nearest neighbor) and temporal (400 ns) proximity. The heat generated by each spiking event transfers to the neighboring neuristors, making their IMT more likely and thus triggering a cascade of spikes. Fig.~\ref{fig:avalanche}(a)(b) shows examples of avalanche size distributions in which a power-law distribution is observed, indicative of LRO. The methodology for identifying these avalanches is detailed in the methods section.

We varied the input voltage and thermal capacitance, $C_\mathrm{th}$, to generate the phase diagram depicted in Fig.~\ref{fig:avalanche}(c). Here, the y-axis reflects $C_\mathrm{th}$'s relative value against the experimentally estimated one. Similar to observations in Fig.~\ref{fig:snapshots}, both a synchronized rigid state, characterized by collective neuristor firing, and a quiescent state, with no spiking activity, are found. Around the phase boundaries, a wide range of parameters leads to a power-law distribution in avalanche sizes across several orders of magnitude, confirming the existence of LRO. This is further supported in Fig.~\ref{fig:avalanche}(d), where we compute avalanche sizes for each point in the parameter space and plot the absolute value of the exponent from the fitted power-law distribution. Areas without a colored box indicate an unsuccessful power-law fit, with the maximum exponent limited to 6 to remove outliers.

\begin{figure*}
    \centering
    \includegraphics[width=1\linewidth]{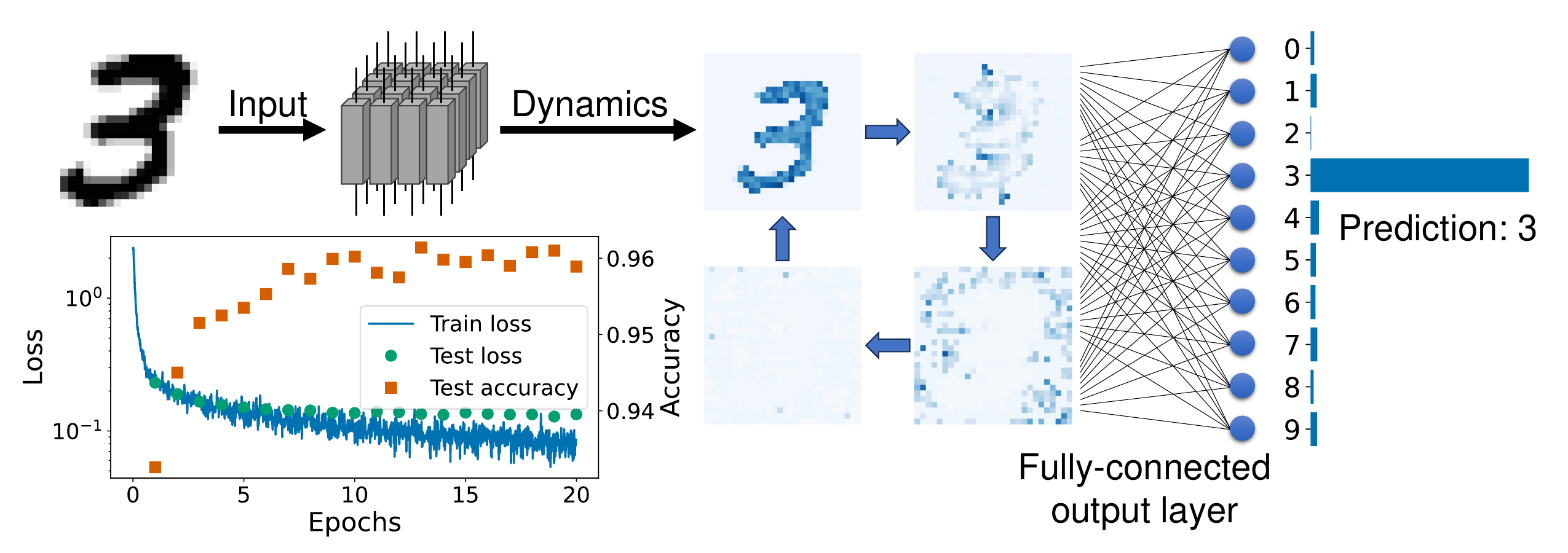}
    \caption{Overview of our reservoir computing implementation with a 2D thermal neuristor array, using the MNIST handwritten digit dataset \cite{lecun1998mnist} as a benchmark. Each image from the dataset is translated into input voltages for a $28\times 28$ thermal neuristor array. The array's spiking dynamics are gathered as the reservoir output. A fully connected output layer, enhanced with softmax nonlinearity, is trained to classify the digit. The bottom-left panel illustrates the training process, displaying both loss and accuracy, culminating in a final test set accuracy of 96\%.}
    \label{fig:MNIST}
\end{figure*}

While we empirically observe power-law scaling in avalanche sizes, one might question if this implies criticality and scale-invariance. The numerical evidence presented here suggests otherwise. First, the power-law distributions in Fig.~\ref{fig:avalanche}(a)(b) do not align with the finite-size scaling ansatz \cite{fisher1972scaling, bonachela2009self}, which predicts diminishing finite-size effects with increasing system size. Furthermore, a rescaling based on the system size should collapse all curves onto one \cite{bonachela2009self, brochini2016phase} for scale-invariant systems, but such an effect is notably missing in our system, contradicting finite-size scaling expectations. In Supplementary Fig.~\ref{fig:collapse}, we present the results of attempted finite-size scaling, which clearly imply a lack of scale-invariance.

Despite the absence of criticality, can the system still perform some computing tasks effectively? Is the LRO observed in these thermal neuristor arrays even necessary for such tasks? We demonstrate in the following section that for classification, LRO, let alone criticality, is not necessary, as anticipated in \cite{cramer2020control}. 

\subsection{Role of LRO in reservoir computing classification tasks}
\label{sec:RC}

We apply our thermal neuristor array to reservoir computing (RC) to answer the above questions. 
RC differentiates itself from traditional neural network models by not requiring the reservoir – the network's core – to be trained. The reservoir is a high-dimensional, nonlinear dynamical system. It takes an input signal, $\mathbf{x}$, and transforms it into an output signal, $\mathbf{y}=f(\mathbf{x})$. A simple output function, usually a fully connected layer, is then trained to map this output signal, $\mathbf{y}$, to the desired output, $\mathbf{\hat{z}}=g(\mathbf{y})$. Training typically involves minimizing a predefined loss function between the predicted output $\mathbf{\hat{z}}$ and the actual label $\mathbf{z}$, associated with the input $\mathbf{x}$, using backpropagation and gradient descent. If the output function is linear, training can be reduced to a single linear regression.

The reservoir's transfer function $f$ can be arbitrary, with its main role being to project the input signal $\mathbf{x}$ into a high-dimensional feature space. Since the reservoir doesn't require training, employing an experimentally designed nonlinear dynamical system like our thermal neuristor array for RC is both effective and straightforward.

As a practical demonstration, we applied RC using thermal neuristors to classify handwritten digits from the MNIST dataset \cite{lecun1998mnist}. Each $28\times 28$ grayscale pixel image, representing digits 0 to 9, is converted into input voltages through a linear transformation. The system is then allowed to evolve for a specific time, during which we capture the spiking dynamics as output features from the reservoir. Subsequently, a fully connected layer with softmax activation is trained to predict the digit. This process is schematically represented in Fig.~\ref{fig:MNIST}.

The output layer was trained over 20 epochs, as shown in the bottom-left panel of Fig.~\ref{fig:MNIST}. The test loss stabilized after approximately 10 epochs, and due to the network's simple architecture, overfitting was avoided. Ultimately, the test set accuracy reached 96\%. Further training details can be found in the methods section.

In this experiment, we treated the voltage transformation, thermal capacitance $C_\mathrm{th}$, and noise strength $\sigma$ as adjustable hyperparameters. This allowed us to check which region of phase space would produce optimal results. We found that the parameters that yielded optimal performance were an input voltage range between 10.5 V and 12.2 V, $C_\mathrm{th}=0.15$, and noise strength $\sigma=0.2$ $\mu$J$\cdot$s$^{-1/2}$. These settings placed us within the synchronized rigid phase, not the LRO one, with the input voltage variations introducing complex oscillatory patterns.  In fact, choosing the parameters in the LRO phase produced worse
results. We show this in Appendix C4 the SI. 
The phase diagram relating to noise strength can be found in Supplementary Fig. 10, and videos of collective oscillations under image inputs are available in Supplementary Movie 2. 

To further explore the role of LRO in reservoir computing tasks, Appendix C5 of the SI details our efforts to eliminate LRO within the reservoir by either removing interactions between neurons altogether or by reducing memory. We quantified LRO using the avalanche size distribution under these various settings. The findings reveal that even when the reservoir operates in a rigid or non-interacting state, long-range structures inherited from the dataset are still apparent. However, no relation between LRO and computational performance was observed. Videos demonstrating collective oscillations under these conditions are available in Supplementary Movies 3 and 4.

As further verification, Appendix C6 of the SI documents an additional experiment involving the prediction of chaotic dynamics governed by the 2D Kuramoto-Sivashinsky equations \cite{kalogirou2015depth}. The results corroborate our primary findings: optimal performance in reservoir computing is achieved without the presence of LRO within the reservoir.

In conclusion, the spiking dynamics of the optimally performing reservoir in our study are not characterized by an LRO state. This observation aligns with the findings in \cite{cramer2020control}, and challenges the well-accepted critical brain hypothesis \cite{hesse2014self} and theories suggesting that near-critical states enhance computational performance \cite{boedecker2012information, del2017criticality}. However, our results do not directly contradict the critical brain hypothesis, since it is possible that long-range correlations are effectively encapsulated within the feed-forward layer. Despite this possibility, our findings highlight a crucial aspect: criticality is not a prerequisite for effective computational performance in such tasks. 

\section{Discussion}
In this study, we have developed and experimentally validated VO$_2$-based thermal neuristors that exhibit brain-like features. We then formulated a theoretical model grounded in our experimental findings to facilitate large-scale numerical simulations. These simulations revealed a variety of phase structures, notably those with LRO, across a broad spectrum of parameters. Our analysis suggests that this LRO stems from the time-nonlocal response of the system and is not associated with criticality. Significantly, we demonstrate that this feature does not impair the system's computational abilities. In fact, it does not even seem to be necessary in some tasks, such as classification and time series prediction, as we have shown by using our thermal neuristor array in reservoir computing. 

The thermal neuristor represents an innovative artificial neuron model, and our research offers insights into the collective dynamics of artificial neuronal activities. Our findings suggest that criticality is not a prerequisite for effective information processing in such systems. This challenges the critical brain hypothesis and its applicability to neuromorphic systems, indicating that even non-critical systems can excel in some computational tasks. We then advocate for a broader exploration of non-critical dynamical regimes that might offer computational capabilities just as powerful, if not more so, than those found at or near a critical state.

Moreover, our work highlights the potential of VO$_2$-based thermal neuristors in computing applications, setting the stage for more extensive experiments. Given the growing need for innovative hardware in neuromorphic computing, our VO$_2$-based thermal neuristor system is a promising candidate for advancing next-generation hardware in artificial intelligence.

\section{Methods}
\label{sec:method}
\subsection{Fabrication of VO$_2$ thermal neuristor arrays}
\subsubsection{Epitaxial VO$_2$ thin film growth}
We employed reactive RF magnetron sputtering to deposit a 100-nm thick VO$_2$ film onto a (012)-oriented Al$_2$O$_3$ substrate. Initially, the substrate was placed in a high vacuum chamber, achieving a base pressure of around $10^{-7}$ Torr, and heated to 680$^\circ$C. The chamber was then infused with pure argon at 2.2 s.c.c.m and a gas mix (20\% oxygen, 80\% argon) at 2.1 s.c.c.m. The sputtering plasma was initiated at a pressure of 4.2 mTorr by applying a forward power of 100 W to the target, corresponding to approximately 240 V. Post-growth, the sample holder was cooled to room temperature at a rate of 12$^\circ$C/min. Specular x-ray diffraction analysis of the film revealed textured growth along the (110) crystallographic direction.

\subsubsection{VO$_2$ thermal neuristor arrays fabrication}
For patterning the VO$_2$ neuristor arrays, Electron Beam Lithography (EBL) was employed. Each neuristor, sized at $100\times 500$ nm$^2$, was delineated with 500 nm gaps. The initial lithography pattern defined electrodes by depositing a 15 nm Ti layer followed by a 40 nm Au layer. To investigate thermal interactions between neuristors, a second lithography and etching step was necessary. We utilized a reactive-ion etching system to etch the exposed VO$_2$ films between devices, as per the second-step lithography patterns, while the negative resist shielded the electrodes and devices from etching.

\subsubsection{Transport measurements}
Transport measurements were conducted in a TTPX Lakeshore probe station equipped with a Keithley 6221 current source, a Keithley 2812 nanovoltmeter, a Tektronix Dual Channel Arbitrary Function Generator 3252C, and a Tektronix Oscilloscope MSO54. The current source and nanovoltmeter were utilized to gauge the device's resistance versus temperature. The Arbitrary Function Generator (AFG) was employed to apply either DC or pulse voltage bursts, while the oscilloscope monitored the output signals. Notably, the impedance for the channel assigned to measure voltage dynamics was set at 1 M$\Omega$, and the channel for capturing spiking current dynamics was configured to 50 $\Omega$.

\subsection{Details of numerical simulations}
\subsubsection{Model details and constant parameters}
\label{sec:const}

The constants in Eqs.~\eqref{eq:current} and \eqref{eq:heat} are crucial in our simulations, as they depend on specific experimental setups. Following the approach in \cite{qiu2024reconfigurable}, we optimized these parameters to closely replicate the experimental results. The chosen values are summarized in Table~\ref{tab:parameters}.

The resistance of the VO$_2$ film, $R$, is modeled based on the hysteresis model introduced in \cite{de2002modeling}, described by the equations:

\begin{equation}
\begin{aligned}
    R(T) &= R_\mathrm{0} \exp\left(\frac{E_\mathrm{a}}{T}\right) F(T) + R_\mathrm{m},\\
    F(T) &= \frac{1}{2} + \frac{1}{2} \tanh \bigg( \beta \bigg\{ \delta\frac{w}{2} + T_\mathrm{c} \\
    & - \left[ T + T_{\mathrm{pr}} P\left( \frac{T - T_\mathrm{r}}{T_{\mathrm{pr}}} \right) \right] \bigg\} \bigg),\\
    T_{\mathrm{pr}} &=\delta\frac{w}{2} + T_\mathrm{c} - \frac{1}{\beta} [2F(T_\mathrm{r}) - 1] - T_\mathrm{r},\\
    P(x) &= \frac{1}{2} \left(1 - \sin \gamma x \right) \left[1 + \tanh \left(\pi^2 - 2\pi x \right) \right].
\end{aligned}\label{eq:hysteresis}
\end{equation}

Each component of Eq.~\eqref{eq:hysteresis} is detailed in \cite{de2002modeling}. The term $T_\mathrm{r}$ denotes the reversal temperature, marking the most recent transition between heating and cooling processes. Here, $\delta$ equals 1 on the heating branch and -1 on the cooling branch, with all other symbols representing constant parameters. These constants were selected in accordance with \cite{qiu2024reconfigurable} to accurately reflect the experimentally observed hysteresis loop, and their values are compiled in Table~\ref{tab:parameters}.

The noise strength $\sigma$ was chosen to facilitate a diverse range of phase structures. We conducted preliminary tests on the phase diagram by varying $\sigma$, with the results detailed in Appendix C3 in the SI.

\begin{table}
    \centering
    \begin{tabular}{l|l|l}
    Param           & Value             & Physical meaning\\\hline
    $C$                 & 145 pF            & Capacitance\\
    $R^\mathrm{load}$   & 12.0 k$\Omega$    & Load resistance\\
    $C_\mathrm{th}$     & 49.6 pJ/K         & Thermal capacitance\\
    & &{\it Note: Figures show relative value to this}\\
    $S_\mathrm{e}$      & 0.201 mW/K        & Thermal conductance to environment\\
    $S_\mathrm{c}$      & 4.11 $\mu$W/K     & Thermal conductance to neighbor\\
    $T_0$               & 325 K             & Ambient temperature\\
    $\sigma$            & 1 $\mu$J$\cdot\mathrm{s}^{-1/2}$          & Noise strength\\
    $R_\mathrm{0}$               & 5.36 m$\Omega$    & Insulating resistance prefactor\\
    $E_\mathrm{a}$               & 5220 K            & VO$_2$ activation energy\\
    $R_\mathrm{m}$               & 1286 $\Omega$     & Metallic resistance\\
    $w$                 & 7.19 K            & Width of the hysteresis loop\\
    $T_\mathrm{c}$               & 332.8 K           & Center of the hysteresis loop\\
    $\beta$             & 0.253             & Fitting parameter in  hysteresis\\
    $\gamma$            & 0.956             & Fitting parameter in  hysteresis\\
    \end{tabular}
    \caption{Parameters utilized in the numerical simulations for Eqs.~\eqref{eq:current}-\eqref{eq:hysteresis}. $R^\mathrm{load}$ and $T_0$ are taken from experiment, while other
    parameters are optimized to align the numerical model as closely as possible to the experimentally measured data. }
    \label{tab:parameters}
\end{table}

\subsubsection{Numerical methods}
\label{sec:numerical_methods}

For the numerical integration of Eqs.~\eqref{eq:current} and \eqref{eq:heat}, we employed the Euler-Maruyama method \cite{maruyama1954transition} with a fixed time step of $dt=10$ ns. The current-time trajectories were recorded, and current spikes were identified by locating the local maxima within these trajectories.

To analyze the avalanche size distribution, we first defined an ``avalanche'' as a contiguous series of spiking events occurring within a certain spatial and temporal proximity. We determined a specific window length for both spatial and temporal dimensions and then coarse-grained the spiking trajectories, categorizing each spiking event into a corresponding window. This process resulted in a $D+1$-dimensional lattice ($D$ spatial and 1 temporal dimensions), where each lattice site denoted the number of spikes within its window. Following this, the Hoshen-Kopelman algorithm \cite{hoshen1976percolation} was applied to identify clusters of spiking activities within the lattice. Each identified cluster was considered as one distinct avalanche, in line with our defined criteria.

The avalanche size distribution is influenced by the chosen window size. Generally, the temporal window length should be significantly longer than the duration of each spike but shorter than the interval between consecutive spikes. For all results presented in this paper, the temporal window length was set at 400 ns. In terms of spatial window length, we focused on immediate neighbors (length = 1) of each neuristor for cluster identification.

After identifying the avalanches, we computed the histogram of avalanche sizes using a logarithmic binning scheme \cite{pruessner2012self}, where bins are uniformly distributed on a logarithmic scale. The sizes of these bins were determined according to Scott's normal reference rule \cite{scott1979optimal}. To characterize the avalanche size distributions presented in Fig.~\ref{fig:avalanche}, we applied a power-law fit to each histogram, excluding the tails for more accurate modeling.

\subsection{Reservoir setup}
\label{sec:reservoir}

In employing thermal neuristors for reservoir computing, we consider the entire neuristor array as the reservoir. The input voltages serve as the reservoir input, and the resultant spike trains are recorded as the output.

For the MNIST dataset \cite{lecun1998mnist}, the reservoir's parameters are detailed in the main text. To record the spike trains, we simulate the system dynamics for 10 $\mu$s, extracting spikes using the method outlined in the previous section. These spike trains are then coarse-grained with a time window of $\Delta t=500$ ns. Each time window is assigned a binary value indicating the presence or absence of a spike. This process results in a $28\times 28\times 20$ binary array representing the reservoir's spike train output. This array is then flattened into a one-dimensional sequence of 15680 elements. A fully connected layer with dimensions $15680\times 10$ is trained to map the reservoir output to the ten digit classes. At the final stage, a softmax nonlinearity is applied to transform the output layer's results into predicted probabilities. Although activation functions are not typically standard in reservoir computing tasks, we still implemented the softmax activation in conjunction with negative log-likelihood loss, as it demonstrated enhanced performance compared to mean-square-error loss without an activation function.

For training this fully connected output layer, we utilized the Adam optimizer \cite{kingma2014adam} with a learning rate of $10^{-3}$. The corresponding loss curve is depicted in the bottom-left panel of Fig.~\ref{fig:MNIST}.

\section*{Data availability}
The MNIST handwritten digit dataset can be accessed at \cite{lecun1998mnist}. The experimental data have been deposited in the github repository \cite{github_link}. All other data that support the findings of this study can be generated by running the code at \cite{github_link}.

\section*{Code availability}
A demo code to reproduce the results in this work can be found at \cite{github_link}.

\section*{Acknowledgments}
Y.-H.Z., and M.D. were supported by the Department of Energy under Grant No. DE-SC0020892. C.S. was supported by the Center for Memory and Recording Research at UCSD. E.Q. and I.K.S. were supported by the Air Force Office of Scientific Research under award number FA9550-22-1-0135.  

\section*{Author contributions}
Y.-H.Z. and M.D. conceived the project. M.D. supervised the theoretical work. Y.-H.Z. did all the numerical calculations. C.S. performed the analysis on the emergence of LRO from memory. E.Q. and I.K.S. provided the experimental data. All authors have read and contributed to the writing of the paper.

\section*{Competing Interests}
The authors declare no competing interests.

\bibliography{bibliography}

\clearpage
\newpage
\appendix

\begin{center} 
{\large \bf Supplementary Information: Collective dynamics and long-range order in thermal neuristor networks}
\end{center} 

\setcounter{page}{1}
% \setcounter{figure}{0}
% \setcounter{equation}{0} 
% \makeatletter
% \renewcommand{\thefigure}{S\arabic{figure}}
% \renewcommand \theequation{S\@arabic\c@equation}
% \renewcommand \thetable{S\@arabic\c@table}
% \setcounter{secnumdepth}{3}
% \makeatother

\section{Emergence of long-range order}

In this section, we delve deeper into Eqs.~\eqref{eq:current} and \eqref{eq:heat} to explore the emergence of long-range order (LRO). In particular, we show that it arises due to the time non-local (memory) response of the system.

Heuristically, the relatively slowly varying temperature field gradually couples spatially separated neuristors, giving rise to long-range correlations. This intuition suggests that only terms in these equations that couple $V_i$ and $T_i$ or introduce thermal diffusive coupling are necessary to induce LRO. As our analysis continues, it will become more clear that these are, in fact, the only terms that are necessary to make the existence of LRO manifest. For clarity, we rewrite Eqs.~\eqref{eq:current} and \eqref{eq:heat} below, with ``$\dots$'' representing the terms in the original equations which are now irrelevant for the purposes of our analysis:

\begin{equation}
\begin{split}
    \dot{V}_i & = \frac{- V_i}{\tau_{V_i}} + \dots\,, \\
    \dot{T}_i & = \frac{1}{\tau_{\text{T}}} \bigg( \frac{V_i^2}{S_{\text{c}} R_i} + \nabla^2 T_i \bigg) + \dots \,. \label{eq:simplifiedeqns}
\end{split}
\end{equation}

Here, $\tau_{V_i} \equiv C R_i$ represents a site-dependent voltage timescale, and $\tau_{\text{T}} \equiv C_{\text{th}}/S_{\text{c}} \approx 50 \tau_{\text{th}}$ is the intersite thermal diffusion timescale. The latter differs from $\tau_{\text{th}}$ by a factor of about 50, as the thermal coupling between sites is significantly weaker than with the environment. To simplify our analysis, we assume $\tau_{V_i}^{-1}$ is a linear interpolation between the inverse timescale in the insulating state ($\tau_{\text{ins}}^{-1}$) and the metallic state ($\tau_{\text{met}}^{-1}$), across a range within the hysteresis loop (see Fig.~\ref{fig:schematic}(b) in the main text), governed by the following equations: 

\begin{equation}
    \frac{1}{\tau_{V_i}} = \frac{\alpha(T_i)}{\tau_{\text{ins}}} + \frac{1 -\alpha(T_i)}{\tau_\text{met}}\,,
\end{equation}
%\vspace{}
\begin{equation}
    \alpha(T_i) = \frac{1}{2} \Bigg( 1 - \bigg( \frac{T_i - T_{\text{c}}}{w} \bigg) \Bigg)\,.
\end{equation}

Here, $T_{\text{c}}$ and $w$ characterize the center and width of the hysteresis loop, respectively.

This approximation, which defines $\alpha(T_i)$, is inspired by Matthiessen's rule \cite{Ashcroft76}, traditionally applied in scattering contexts. By focusing solely on $T_i$-dependent terms in $\tau_{V_i}^{-1}$, we explicitly couple the current and thermal dynamics without reference to $R_i$:

\begin{equation}
\begin{split}
    \dot{V}_i & = \frac{1}{\tau_V} \bigg(\frac{-T_i V_i}{T_{\text{c}}} \bigg) + \dots\,, \\
    \dot{T}_i & = \frac{1}{\tau_T} \bigg( \zeta T_i V_i^2 + \nabla^2 T_i \bigg) + \dots\,. \label{eq:finaleqns}
\end{split}
\end{equation}

In \eqref{eq:finaleqns}, $\tau_V^{-1} \equiv \frac{T_{\text{c}} (\tau_{\text{ins}} - \tau_{\text{met}})}{2 w \tau_{\text{ins}} \tau_{\text{met}}} \approx \frac{T_{\text{c}}}{2 w \tau_{\text{met}}}$ (since $\tau_{\text{ins}} \gg \tau_{\text{met}}$) and $\zeta \equiv \frac{C}{2 w \tau_{\text{met}} S_{\text{c}}}$. Additionally, we confirm that both $T_i/T_{\text{c}} \sim 1$ and $\zeta V_i^2 \sim 1$ with our chosen parameter values (see Table~\ref{tab:parameters}, and note that $V_i \sim 5 \text{V}$ during spiking dynamics). Therefore, $\tau_V$ and $\tau_T$ are representative \textit{site-independent} timescales of the relevant current and thermal dynamics, respectively. Given our chosen parameter values, $\tau_V\sim$ 10 ns and $\tau_T\sim$ 10 $\mu$s, indicating a separation of timescales.

Considering these characteristic timescales $\tau_V$ and $\tau_T$, we can treat the contributions to $\dot{T}_i$ written in \eqref{eq:finaleqns} as small over intervals $\Delta t \lesssim \tau_T$. Thus, the relevant contributions to $T_i$ over each such interval will be approximately constant. Although experiment/simulation suggests a more comprehensive thermal timescale might be shorter (the inter-spike interval is $\sim 1 \mu\text{s}$), recall that we are only considering terms that couple $T_i$ and $V_i$ or introduce diffusive coupling. While $T_i$ does implicitly depend on voltage coupling, thermal coupling with adjacent sites and the environment, \textit{and} noise, contributions to $T_i$ which only depend on the time-evolution terms kept in Eq.$~\eqref{eq:finaleqns}$ will always exist. For clarity, we call these relevant contributions $\tilde T_i$ and $\tilde V_i$.

We now show that by iteratively integrating $\tilde T_i$ and $\tilde V_i$ over a prolonged period of time, the presence of long-range current couplings becomes manifest. This ``time coarse-graining'' approach is analogous to methods used in other complex systems studies \cite{wilson_cowen, kirkwood}.

First, we evolve $\tilde T_i$ over intervals of length $\tau_T$, during which $\tilde T_i$ is approximately constant:

\begin{equation}
\begin{split}
    \tilde T_i(\tau_T) & = T_i(0) + \int_0^{\tau_T} \dot{\tilde T}_i(t) dt \\
    & \approx \big(1 + \zeta \overline{V_{i, 1}^2} + \nabla^2 \big) T_i(0) \equiv \hat{L}_0 T_i(0)\,,
\end{split}
\end{equation}

\begin{equation}
\begin{split}
    \tilde T_i(2 \tau_T) & = \tilde T_i(\tau_T) + \int_{\tau_T}^{2 \tau_T} \dot{\tilde T}_i(t) dt \\
    & \approx \big(1 + \zeta \overline{V_{i, 2}^2} + \nabla^2 \big) \tilde T_i(\tau_T) \equiv \hat{L_1} \hat{L_0} T_i(0)\,.
\end{split}
\end{equation}

Above, we've defined $\overline{V_{i, l}^2}$ to be the average of $V_i^2$ over a time interval $[(l-1)\tau_T, l\tau_T)$ (since $\tau_V << \tau_T$, $V_i^2$ is not approximately constant over intervals of length $\tau_T$, and there is no need to distinguish between $V_i$ and $\tilde V_i$). We introduce the operator $\hat{L}_p$, which time-evolves $\tilde T_i(p \tau_T)$ to $\tilde T_i((p + 1) \tau_T)$, for compactness. This generalizes to

\begin{equation}
\begin{split}
    \tilde T_i(l \tau_T) & \approx \Bigg( \prod_{p = 0}^{l - 1} \hat{L}_p \Bigg) T_i(0)\,. \label{eq:coarsegrainthermal_l}
\end{split}
\end{equation}

For $l \geq 2$, notice that $\tilde T(l\tau_T)$ will have terms $\sim \nabla^2 \overline{V_{i, l-1}^2}$ to $\sim \nabla^2 \dots \nabla^2 \overline{V_{i, 1}^2}$, the latter of which has $l-1$ $\nabla^2$'s. These terms implicitly depend on $V_j$ through the diffusive coupling, where $j$ can range from the 1$^\text{st}$ to $(l-1)^\text{th}$ nearest-neighbor of the $i^\text{th}$ site. Applying a similar technique for $\tilde V_i$,

\begin{equation}
\begin{split}
    \tilde V_i(\tau_T) &= V_i(0) + \int_0^{\tau_T} \dot{\tilde V}_i(t) dt \\ & \approx V_i(0) - \frac{\tau_T}{\tau_V} \frac{1}{T_{\text{c}}} \big( T_i(0) \overline{V_{i, 1}^2} \big)\,,
\end{split}
\end{equation}

\begin{equation}
\begin{split}
    \tilde V_i(2\tau_T) &\approx \tilde V_i(\tau_T) - \frac{\tau_T}{\tau_V} \frac{1}{T_{\text{c}}} \big( \tilde T_i(\tau_T) \overline{V_{i, 2}^2} \big) \\
    & \approx V_i(0) - \frac{\tau_T}{\tau_V} \frac{1}{T_{\text{c}}} \big( T_i(0) \overline{V_{i, 1}^2} + \tilde T_i(\tau_T) \overline{V_{i, 2}^2} \big)\,.
\end{split}
\end{equation}

Again, this generalizes for $\tilde V_i(N \tau_T)$:

\begin{equation}
    \tilde V_i(N \tau_T) \approx V_i(0) - \frac{\tau_T}{\tau_V} \frac{1}{T_{\text{c}}} \sum_{l = 0}^{N - 1} \tilde T_i(l \tau_T) \overline{V_{i, l+1}^2}\,.
\end{equation}

\begin{figure*}
    \centering
    \includegraphics[width=1\linewidth]{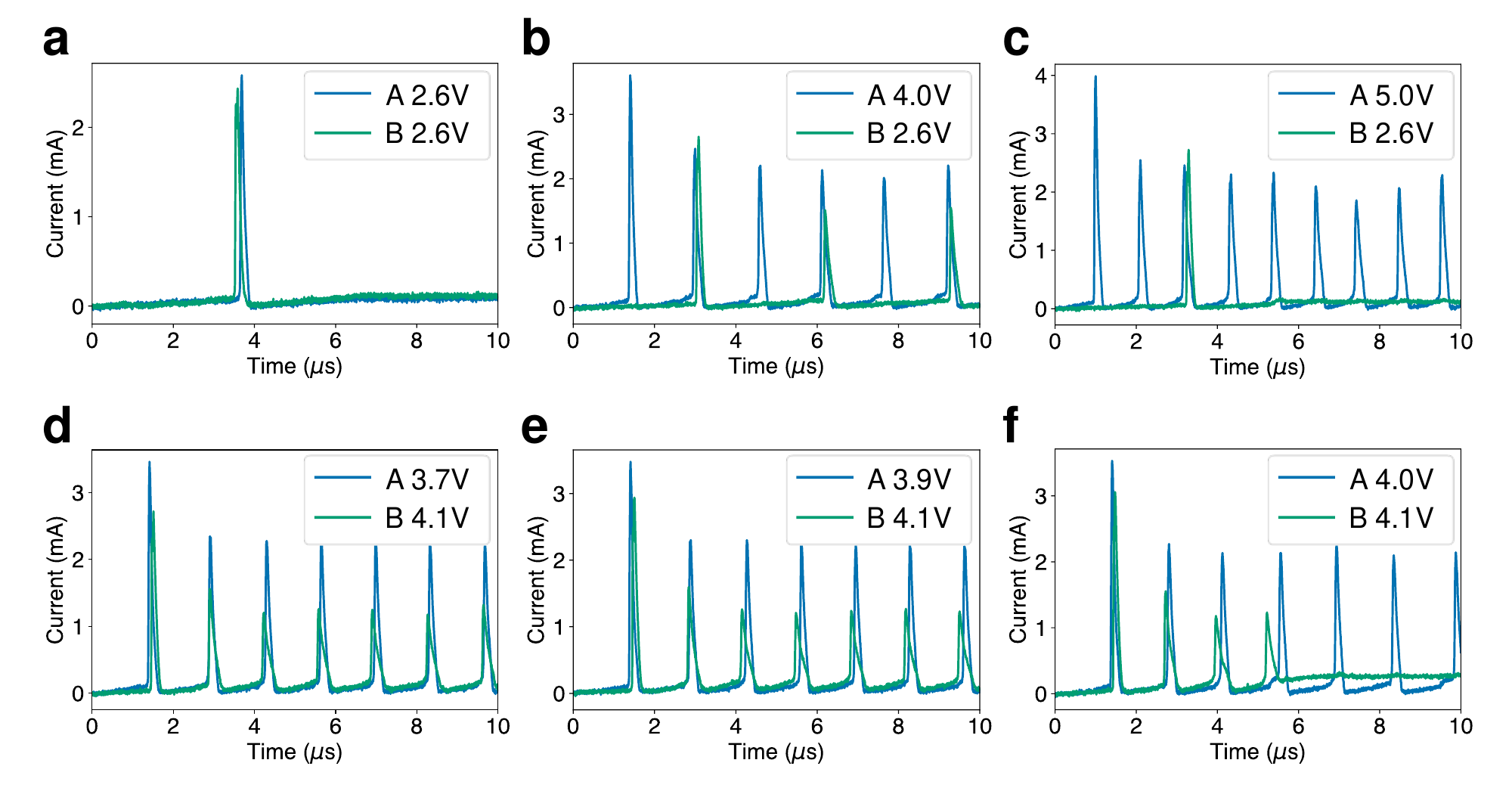}
    \caption{Experimental demonstration of thermal interactions between two adjacent thermal neuristors. Panels (a)-(c) feature neuristor B under a 2.6 V input, just below its lower spiking threshold. Panels (d)-(f) depict neuristor B with a 4.1 V input, marginally below the upper spiking threshold. (a) A 2.6 V input to neuristor A triggers a single synchronized spike in both neuristors. (b) Increasing A's input to 4 V establishes a stable 2:1 spiking pattern. (c) Further increasing A's input to 5 V disrupts the synchronization, resulting in only one spike in B. (d) A 3.7 V input to A leads to 1:1 stable spiking oscillations. (e) A slight increase in A's input voltage introduces phase lags in the synchronization. (f) A further increase in A's voltage breaks the synchronization, causing B to stop spiking.}
    \label{fig:thermal_interaction}
\end{figure*}

As is suggested by Eq.~\eqref{eq:coarsegrainthermal_l}, $\tilde T_i(l \tau_T)$ will implicitly depend on voltages $l - 1$ sites away from the $i^\text{th}$ lattice site. Since there is feedback between $V_i$ and $T_i$, this means that $\tilde V_i(N \tau_T)$ will also depend on other $V_j$ in a highly non-local manner. Additionally, after a sufficiently large time $N \tau_T$, these non-local couplings will have arbitrarily large order (the number of $V_j$ which multiply one another). This spatial non-locality, induced by the interplay between $V_i$-$T_i$ couplings and thermal site diffusion, is indicative of the LRO that we observe numerically.

\section{Experimentally measured thermal interactions}

Here, we provide some additional experimentally measured interactions between two neighboring thermal neuristors. Despite only having two neuristors, the interactions observed here offer valuable insights into the various phases of the thermal neuristor array.

In Fig.~\ref{fig:thermal_interaction}, we show the current-time curves of two adjacent neuristors under varying input voltages. Note that these devices are different from those discussed in the main text, possessing distinct threshold voltages, though their qualitative behaviors are consistent. 

In panels (a) to (c) of Fig.~\ref{fig:thermal_interaction}, neuristor B is set with a 2.6 V input voltage, slightly below the threshold for independent spiking. However, the influence of neuristor A initiates spiking activities. Panel (a) shows that a 2.6 V input to neuristor A, also below its spiking threshold, results in mutual heating and a single synchronized spike in both neuristors. Increasing A's driving voltage leads to a stable spike train in A, as seen in panel (b), which in turn induces stable oscillations in B, forming a 2:1 spiking pattern. Further increasing A's voltage disrupts this synchronization, resulting in only a single spike in B, as depicted in panel (c).

Conversely, in panels (d) to (f), neuristor B operates under a 4.1 V input, capable of sustaining stable oscillations alone but sensitive to additional stimuli. Panel (d) shows that an optimal input to A yields a 1:1 synchronized pattern. However, a slight increase in A's voltage, as shown in panel (e), introduces phase lags between A and B. Eventually, as illustrated in panel (f), the significant phase lag disrupts B's stable spiking pattern, causing it to cease spiking.

In this example, both boundaries of the spiking thresholds exhibit first-order phase transitions, where minor changes in external stimuli lead to markedly different behaviors. This underpins the phase boundaries identified in Fig.~\ref{fig:avalanche} in the main text. Slight perturbations are amplified and propagated through the lattice, resulting in the diverse phase structures and LRO we observed.

\section{Additional numerical results}

\subsection{Spiking patterns and avalanche size distributions under different conditions}

In this section, we provide an intuitive understanding and visualization of spiking patterns and their corresponding avalanche size distributions under various conditions.  The settings used here are consistent with those in Fig.~\ref{fig:snapshots} and Fig.~\ref{fig:avalanche}, and the figures presented retain the same meanings as those referenced.

\begin{figure}
    \centering
    \includegraphics[width=1\linewidth]{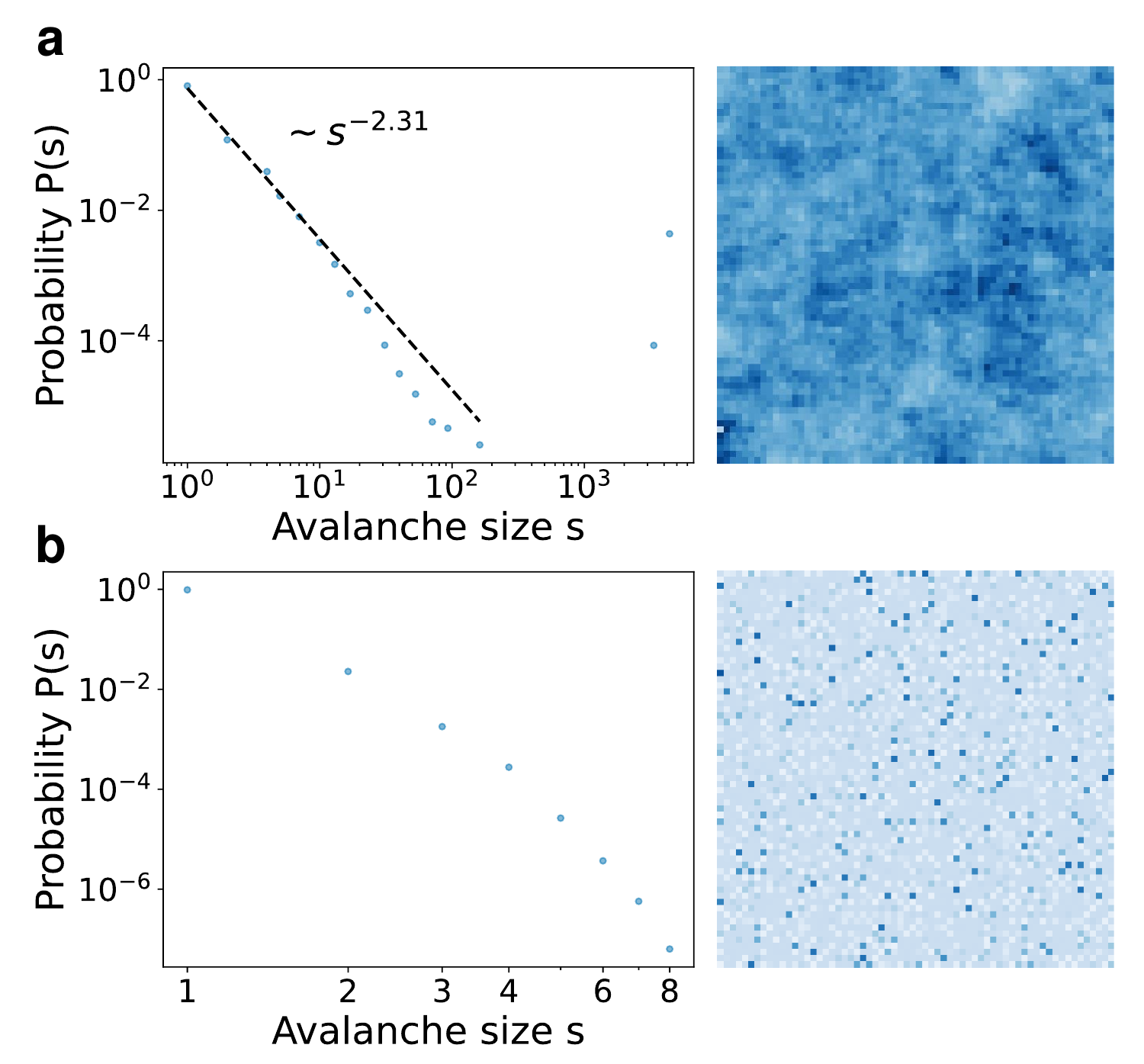}
    \caption{Examples of avalanche size distributions and corresponding spiking patterns for settings consistent with Fig.\ref{fig:snapshots} and Fig.\ref{fig:avalanche} in the main text: (a) $V^{\mathrm{in}}=12.5$ V, $C_{\mathrm{th}}=1$. This setting corresponds to the rigid phase as depicted in Fig.~\ref{fig:avalanche}(c). The right panel illustrates synchronized spiking behavior across nearly all neuristors, with the avalanche size distribution showing a power-law for smaller events. Notably, a prominent peak at the right end of the distribution indicates the prevalence of system-wide avalanches.  (b) $V^{\mathrm{in}}=13$ V, $C_{\mathrm{th}}=1.25$. This setting is at the boundary of the ``no activity'' phase. The snapshot reveals random, uncorrelated spikes, and the avalanche size distribution is predominantly characterized by single or very small events.}
    \label{fig:example}
\end{figure}

First, to elucidate the phase structures depicted in Fig.~\ref{fig:avalanche}, Fig.~\ref{fig:example} illustrates the avalanche size distributions and corresponding spiking patterns at a point within the rigid phase ($V^{\mathrm{in}}=12.5$ V, $C_{\mathrm{th}}=1$) and at a point near the no activity phase ($V^{\mathrm{in}}=13$ V, $C_{\mathrm{th}}=1.25$).

\begin{figure}
    \centering
    \includegraphics[width=1\linewidth]{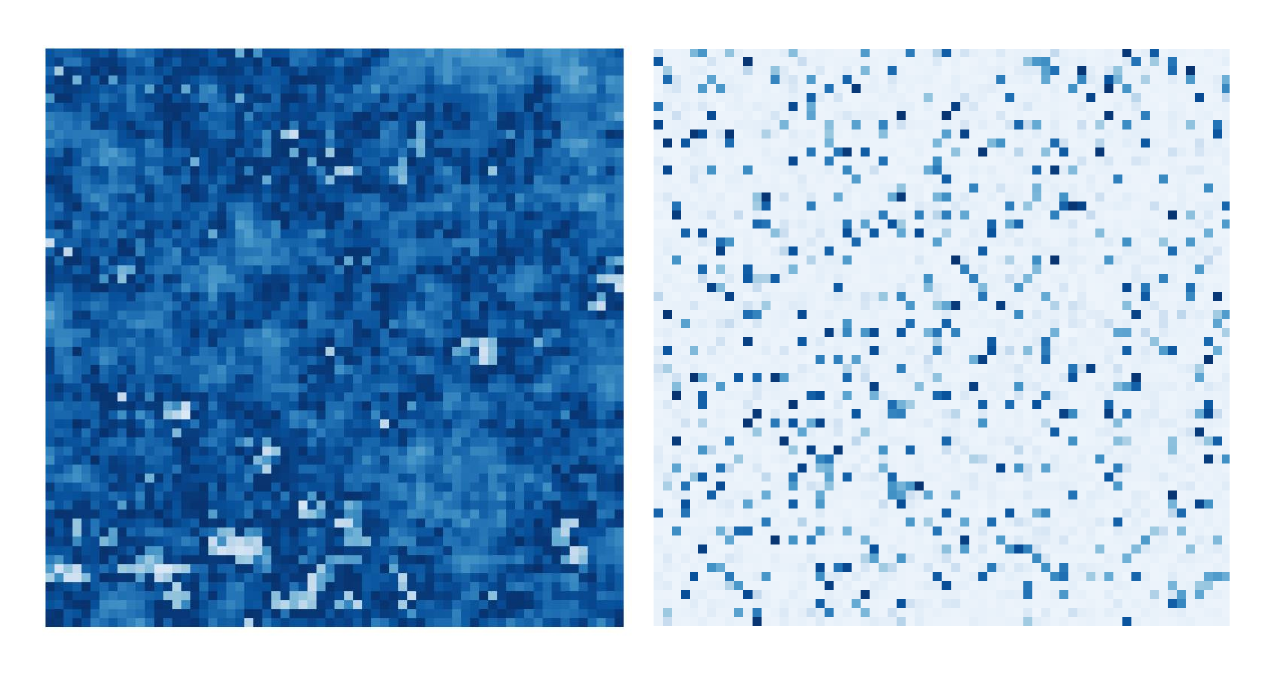}
    \caption{Comparison of spiking patterns with (left) and without (right) thermal coupling at $V^{\mathrm{in}}=12$ V, $C_{\mathrm{th}}=1$. With thermal coupling, the setting induces a rigid state where almost all neuristors spike simultaneously, demonstrating highly correlated behavior. Conversely, without thermal coupling, the neuristors spike independently, resulting in uncorrelated, isolated spiking patterns.}
    \label{fig:couple}
\end{figure}

Furthermore, to highlight the significance of thermal coupling, Fig.~\ref{fig:couple} compares spiking patterns with and without thermal coupling at $V^{\mathrm{in}}=12$ V and $C_{\mathrm{th}}=1$. Given that thermal coupling is the only means through which neuristors exchange information, the absence of thermal coupling naturally results in random, uncorrelated spikes.

\begin{figure}
    \centering
    \includegraphics[width=1\linewidth]{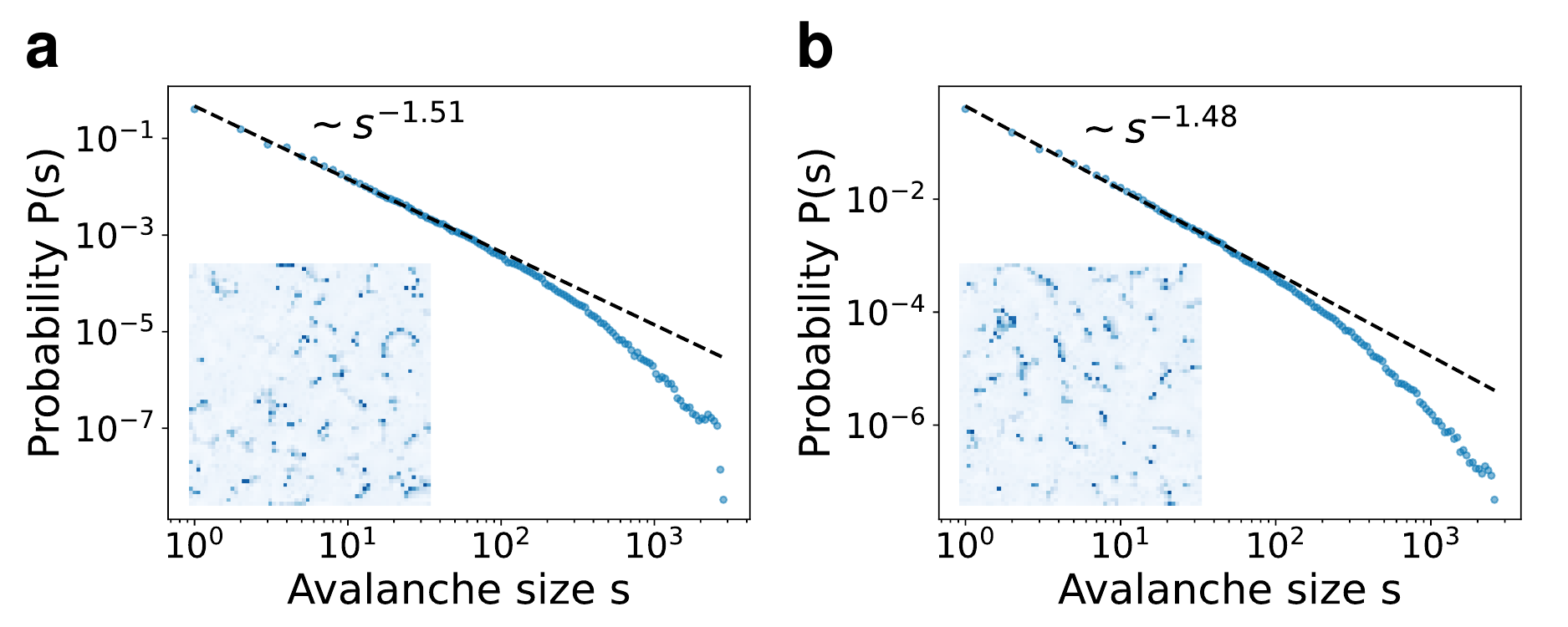}
    \caption{Comparison of spiking patterns and avalanche size distributions under (a) periodic and (b) open boundary conditions at $V^{\mathrm{in}}=9.96$ V and $C_{\mathrm{th}}=1$, both settings within the LRO phase. The figure demonstrates that the type of boundary condition has minimal impact on the spiking behaviors and avalanche size distributions, indicating robustness of the spiking dynamics against boundary effects.}
    \label{fig:PBC}
\end{figure}

Additionally, in Fig.~\ref{fig:PBC}, we compare the spiking patterns and avalanche size distributions under periodic and open boundary conditions at $V^{\mathrm{in}}=9.96$ V and $C_{\mathrm{th}}=1$. This comparison shows that the type of boundary condition has minimal impact on the spiking behaviors, demonstrating the robustness of the spiking dynamics against boundary effects.

\subsection{Finite-size scaling}

\begin{figure*}
    \centering
    \includegraphics[width=1\linewidth]{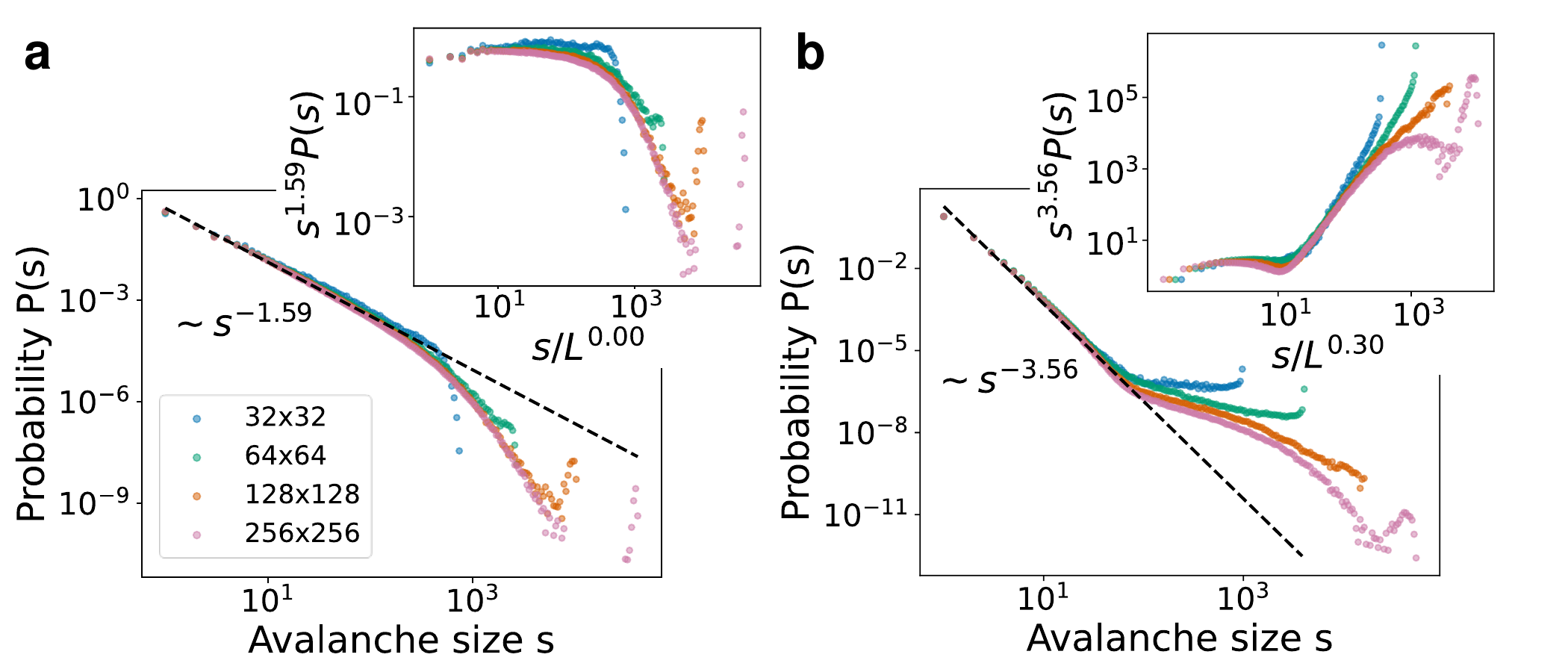}
    \caption{Rescaling of the avalanche size distributions from Fig.~\ref{fig:avalanche} in the main text, based on the finite-size scaling ansatz, Eq.~\eqref{eq:finite_size}. In the inset, $L=\sqrt{N}$ is the length of the lattice. Criticality should result in an overlap of curves from different system sizes, demonstrating scale-invariance. However, this scale-invariance is notably absent in our system, suggesting a lack of criticality.}
    \label{fig:collapse}
\end{figure*}

In the main text (Fig.~\ref{fig:avalanche}), we observed power-law distributions of avalanches across various system sizes, with closely matched power-law exponents at specific parameter points.

According to finite-size scaling theory \cite{fisher1972scaling}, at criticality, avalanche size distributions for different system sizes should conform to a common scaling rule:

\begin{equation}
    P(s, N) \sim s^{\alpha} \exp{\left(-s/N^\beta\right)}, \label{eq:finite_size}
\end{equation}
where $s$ is the avalanche size, $N$ is the system size, $\alpha$ is the critical exponent of the avalanche size distribution, and $\beta$ is the cutoff exponent.As emphasized in several studies, such as \cite{brochini2016phase, bonachela2009self}, the rescaling of avalanche size distributions according to Eq.~\eqref{eq:finite_size} should result in all distributions collapsing onto a single curve at criticality, characterized by a near-perfect overlap of all rescaled curves. Indeed, this phenomenon is typical in scale-free systems, with examples of near-perfect rescaling plots documented in references like \cite{brochini2016phase, bonachela2009self, sipling2024memory}.

Fig.~\ref{fig:collapse} attempts this rescaling for the distributions presented in Fig.~\ref{fig:avalanche} from the main text. Here, we optimized the exponent $\beta$ to align the four curves corresponding to different system sizes. However, our findings show a clear departure from Eq.~\eqref{eq:finite_size}, suggesting the absence of scale-invariance and criticality.

\subsection{Effect of noise strength}

\begin{figure*}
    \centering
    \includegraphics[width=1\linewidth]{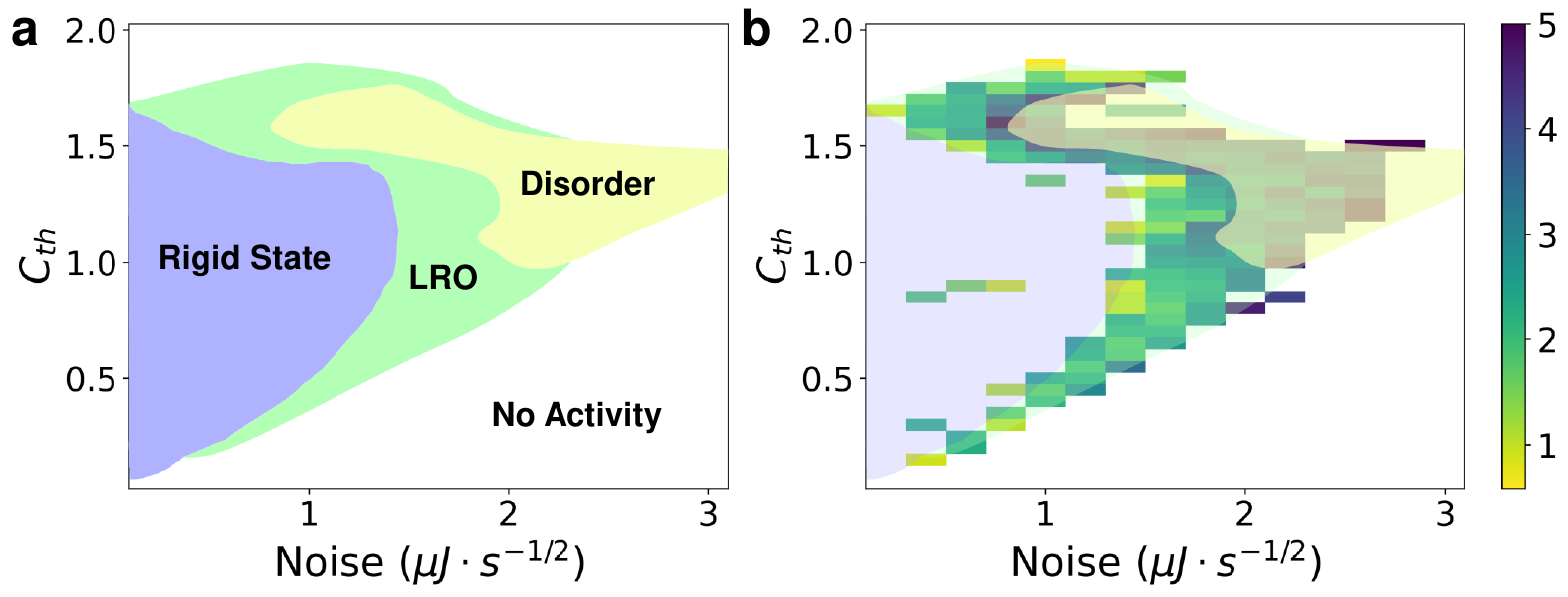}
    \caption{Phase diagram of a $32\times 32$ thermal neuristor array under varying noise strength and thermal capacitance, with a fixed input voltage of 12 V. As noise strength approaches zero, a rigid state prevails, characterized by synchronized firing of all neuristors. With increasing noise, a phase of LRO emerges, eventually giving way to a disordered phase at higher noise levels, characterized by uncorrelated spiking. (b) The slope of the avalanche size distributions at each parameter point, capped at 5 and excluding the negative sign for clarity. A lack of color in a box signals an unsuccessful power-law fit. The phase diagram from panel (a) is overlaid for context. 
    }
    \label{fig:noise}
\end{figure*}

In the main text, we set the noise strength $\sigma$ in Eq.~\eqref{eq:heat} to 1 $\mu$J$\cdot\mathrm{s}^{-1/2}$. This section examines how varying the noise strength influences the phase structures of the system.

Fig.~\ref{fig:noise}(a) presents the phase diagram for a $32\times 32$ thermal neuristor array, exploring the interplay between thermal capacitance and noise strength. Here, the input voltage is consistently held at 12 V, while all other parameters remain as described in the main text. Fig.~\ref{fig:noise}(b) plots the slopes of the avalanche size distributions for each set of parameters, paralleling the approach of Fig.~\ref{fig:avalanche} from the main text. 

It is evident that noise strength significantly impacts the phase structures in this model. Near zero noise strength, when all neuristors are identical and begin with the same initial conditions, a rigid phase emerges across most parameters. This phase is characterized by synchronized spiking in all neuristors. Increasing the noise strength reveals a phase of LRO, situated between the rigid phase and the inactive phase. As the noise strength further escalates, the rigid phase vanishes, giving way to a disordered phase. In this phase, correlations between neuristors fade exponentially, resulting in isolated, uncorrelated spikes. In our main text simulations, we chose a noise strength of 1 $\mu$J$\cdot\mathrm{s}^{-1/2}$, leading to a wide variety of oscillation patterns.

\subsection{Hyperparameter selection in reservoir computing}

\begin{figure*}
    \centering
    \includegraphics[width=1\linewidth]{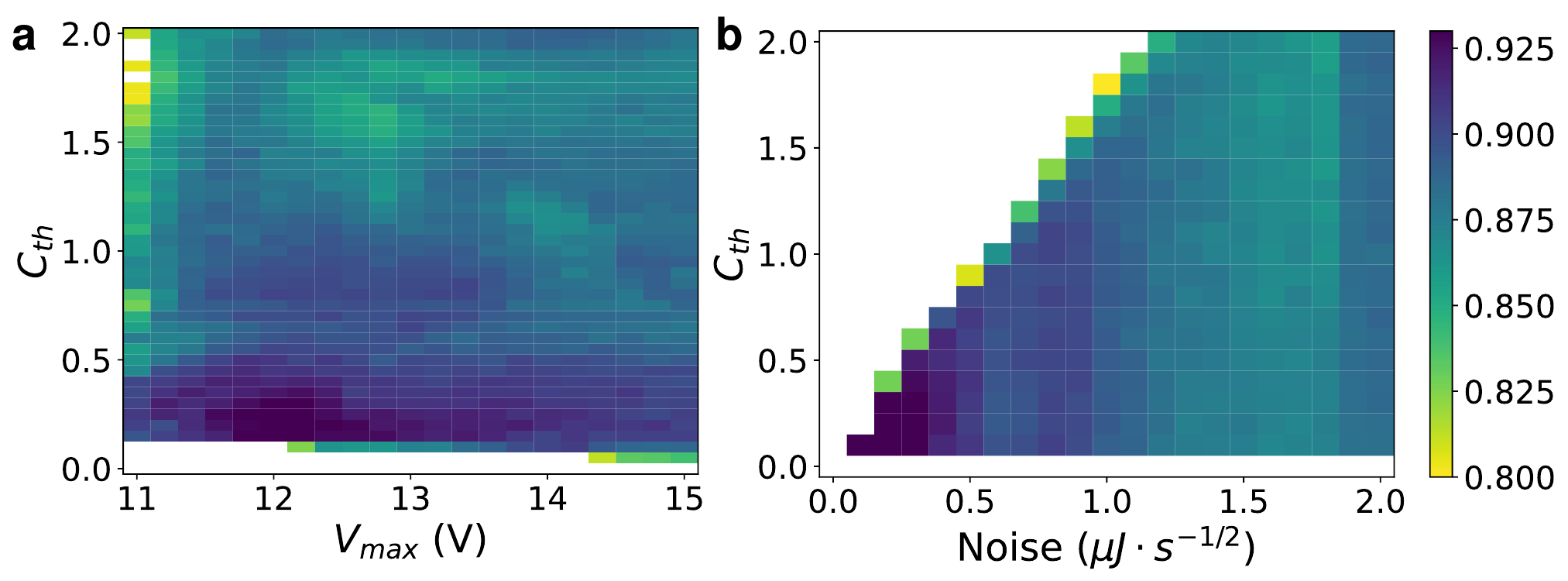}
    \caption{Classification accuracy on the MNIST dataset after one epoch of training with various parameters. Areas in white indicate regions lacking spiking dynamics. (a) Varying the maximum input voltage $V_\mathrm{max}$ and thermal capacitance $C_\mathrm{th}$, while fixing the minimum input voltage $V_\mathrm{min}$ to be 10.5 V and the noise strength $\sigma$ to be 0.2 $\mu$J$\cdot\mathrm{s}^{-1/2}$. Optimal performance is observed at the point $C_\mathrm{th}=0.15$, $V_\mathrm{max}=12.2$ V. (b) Fixing $V_\mathrm{min}=10.5$ V and $V_\mathrm{max}=12$ V. Optimal performance is observed at the point $\sigma=0.2$ $\mu$J$\cdot\mathrm{s}^{-1/2}$ and $C_\mathrm{th}=0.1$. 
    }
    \label{fig:accuracy}
\end{figure*}

Hyperparameter tuning is a critical aspect of machine learning algorithms. In our study, the parameters of the thermal neuristor array serve as hyperparameters for the reservoir computing algorithm.

As detailed in the main text, we use a $28\times 28$ array of thermal neuristors for handwritten digit recognition on the MNIST dataset. While Bayesian optimization algorithms can efficiently select hyperparameters \cite{bergstra2013making}, we opt for a grid search to gain deeper insight into the reservoir's physics. Fig.~\ref{fig:accuracy} illustrates the accuracy of predictions after training for one epoch. In panel (a), we adjusted the thermal capacitance $C_\mathrm{th}$ and the input voltage for black pixels ($V_\mathrm{max}$), while keeping the input voltage for white background pixels ($V_\mathrm{min}$) at 10.5 V, and the noise strength $\sigma$ at 0.2 $\mu$J$\cdot\mathrm{s}^{-1/2}$. Similarly, in panel (b), we fixed $V_\mathrm{min}=10.5$ V and $V_\mathrm{max}=12$ V, and varied $\sigma$ and $C_\mathrm{th}$.

In both panels, most parameter combinations yield reasonable performance, provided some level of spiking dynamics is present in the system. The best performance is achieved near the phase transition boundary between no activity and the rigid state, with a small thermal capacitance ($C_\mathrm{th}\sim 0.15$), moderate input voltage ($V_\mathrm{max}\sim 12$ V), and small noise strength ($\sigma \sim 0.2$ $\mu$J$\cdot\mathrm{s}^{-1/2}$). 
A lower thermal capacitance facilitates a quicker response in the thermal neuristor array, while a smaller noise makes the system more predictable, enhancing the classification task. 

% This analysis reveals that neither criticality nor LRO is essential for effective computational performance. Instead, optimal results are obtained in the synchronized rigid state, challenging the critical brain hypothesis.

\subsection{Further analysis of LRO in reservoir computing}

\begin{figure*}
    \centering
    \includegraphics[width=1\linewidth]{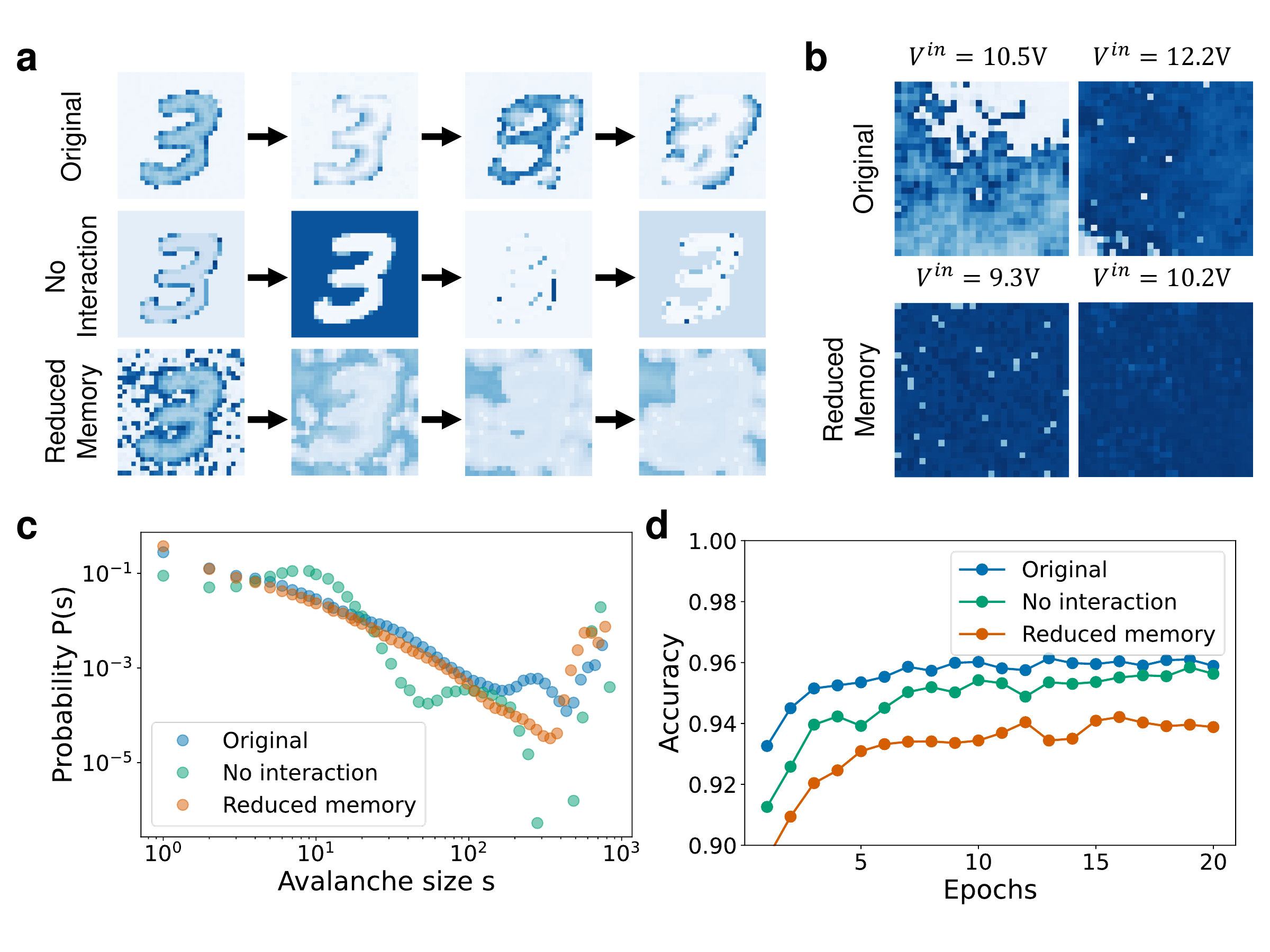}
    \caption{Comparison of MNIST handwritten digit classification using reservoir computing under three different settings. {\bf Original:} Described in the main text with parameters set to  $V_\mathrm{min}=10.50\mathrm{V}$, $V_\mathrm{max}=12.20\mathrm{V}$, $C_\mathrm{th}=0.15$, $\sigma=0.2 \mu\mathrm{J}\cdot\mathrm{s}^{-1/2}$. {\bf No interaction:} Thermal interactions between neighboring neuristors are removed. Parameters are adjusted to $V_\mathrm{min}=11.99\mathrm{V}$, $V_\mathrm{max}=13.37\mathrm{V}$, $C_\mathrm{th}=0.18$, $\sigma=0$, and $S_\mathrm{c}=0$, replacing the original $4.11 \mu$W/K. {\bf Reduced memory:} The slower time scale is minimized by raising the ambient temperature. The parameters are modified to $V_\mathrm{min}=9.30\mathrm{V}$, $V_\mathrm{max}=10.16\mathrm{V}$, $C_\mathrm{th}=0.15$, $\sigma=0.2 \mu\mathrm{J}\cdot\mathrm{s}^{-1/2}$, and $T_0=330$K, up from 325K. Each setting is optimized to achieve the highest possible classification accuracy for the conditions specified. {\bf (a) Snapshots of Reservoir Dynamics:} This panel showcases reservoir dynamics under three distinct settings, with each snapshot taken approximately 4$\mu$s apart. In the original setting, the parameters align the reservoir within the rigid phase, leading the digit pattern to evolve into propagating waves. In the no-interaction scenario, each neuristor's spiking is solely influenced by its individual input voltage, resulting in synchronized spiking across the white background, while pixels corresponding to the digits gradually desynchronize due to minor color variations. Under reduced memory, the reservoir quickly loses detailed information about the digit's shape, leading to a rapid degradation of recognizable patterns. {\bf (b) Dynamics under Uniform Input:} This panel displays reservoir dynamics under uniform input voltages set to $V_\mathrm{min}$ and $V_\mathrm{max}$, confirming that the reservoir operates within a rigid state. The no interaction setting is omitted here, as all neuristors respond identically under uniform voltage conditions. {\bf (c) Avalanche Size Distribution:} Presented here is the avalanche size distribution of current spikes using the MNIST dataset's 60,000 training images as input. In the no-interaction setting, the distribution mirrors the structural characteristics of the dataset, with system-wide avalanches representing the white background and smaller, digit-wide avalanches marking the black pixel areas. With interactions, the distribution for smaller avalanches follows a power-law, smoothing transitions seen in the non-interacting setup, while the configuration of larger avalanches remains intact. {\bf (d) Classification Accuracy:} This panel reports classification accuracy on the MNIST test set after 20 epochs of training. The original configuration, as described in the main text, achieves the highest accuracy at 96.1\%. Remarkably, the no-interaction scenario still performs well, achieving 95.8\% accuracy without any internal information transfer within the reservoir. With reduced memory, where the slower time scale is lessened, long-range order within the system is substantially reduced, leading to a lower accuracy of 94.2\%. These results further illustrate that there is no straightforward correlation between LRO and computational performance.
    }
    \label{fig:MNIST_comparison}
\end{figure*}

\begin{figure*}
    \centering
    \includegraphics[width=1\linewidth]{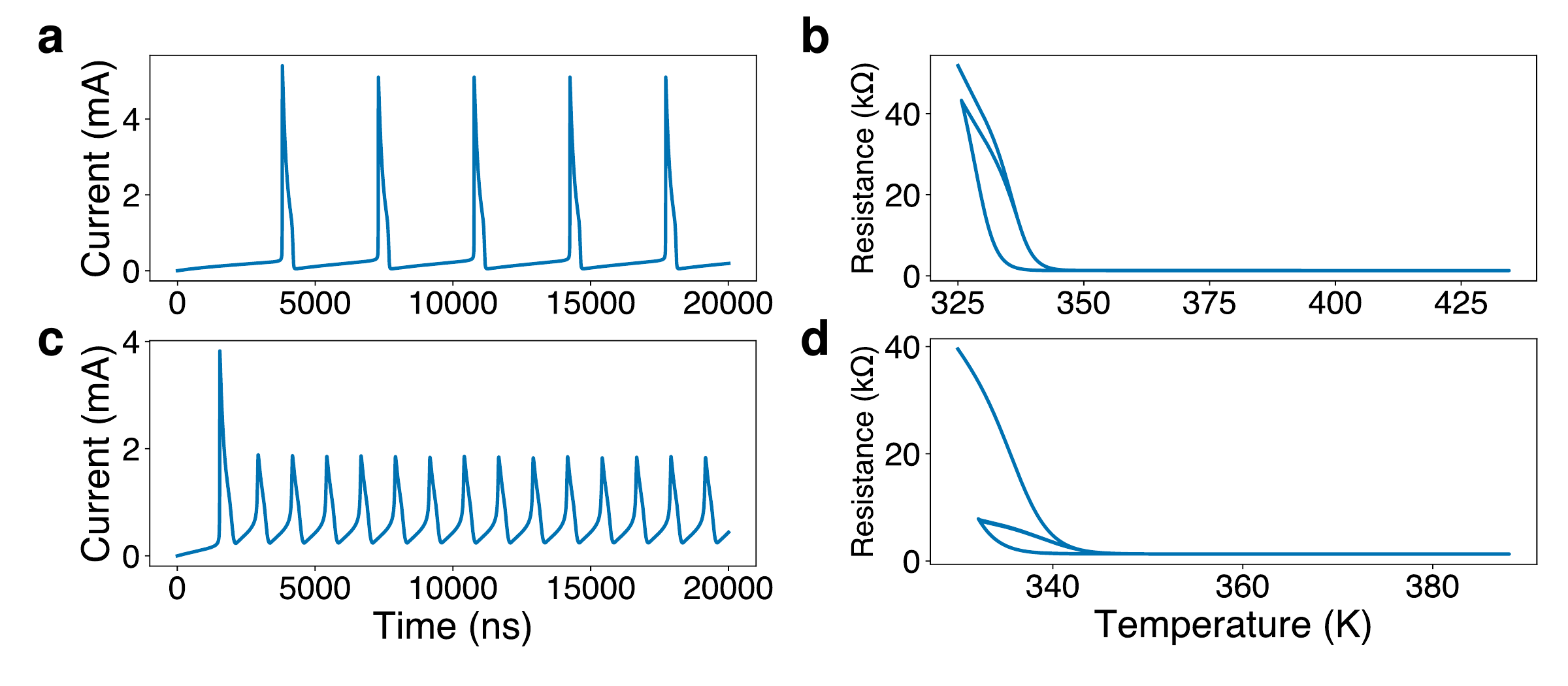}
    \caption{Demonstration of how increasing ambient temperature, $T_0$, reduces the slower time-scale. (a) Spiking dynamics and (b) the resistance-temperature curve of a single neuristor at input voltage $V^\mathrm{in}=10.2$V, $C_\mathrm{th}=0.15$, and $T_0=325$K. The dynamics are stable, with the VO$_2$ device consistently reverting to its insulating state at $R_\mathrm{ins}=43\mathrm{k}\Omega$ after each spike. (c) Spiking dynamics and (d) the resistance-temperature curve for the same neuristor under identical settings but with the ambient temperature increased to $T_0=330$K. The elevated temperature results in the VO$_2$ device reverting to a much lower resistance of 7k$\Omega$ after each spike, significantly reducing the insulating $RC$ time, $\tau_\mathrm{ins}$, to approximately 1$\mu$s. This adjustment leads to smaller spiking amplitudes, a higher frequency, and more unstable spiking dynamics. Further increases in temperature or minor external perturbations can cause the VO$_2$ device to remain in its metallic state, ceasing to spike.
    }
    \label{fig:timescale}
\end{figure*}

In the preceding section, we determined that the optimal parameters place us within the synchronized rigid phase, as depicted in Fig.~\ref{fig:noise}. However, this configuration does not necessarily imply that the reservoir operates in a rigid state, given that variations in the input data may introduce additional structures and correlations within the reservoir.

To rigorously test this hypothesis and explicitly quantify LRO within the reservoir, we computed the avalanche size distribution when the MNIST dataset served as the input. The experimental setup adheres to the methodology outlined in Sec.~\ref{sec:RC} of the main text, employing identical parameters. We recorded the current spikes to calculate the avalanche size distribution using the approach detailed in Sec.~\ref{sec:numerical_methods}. 

The avalanche size distribution, encompassing all 60,000 training images from the MNIST dataset, is depicted in Fig.~\ref{fig:MNIST_comparison}(c). Although the system is in a rigid state, the distribution reveals long-range structures: a power-law distribution is evident for smaller clusters up to $s\sim 10^2$, followed by two prominent bumps, which correspond to avalanches that span across the entire system and those limited to the black pixels comprising the digits.

Superficially, the distinct structure observed within the MNIST dataset’s avalanche size distribution may seem to suggest that LRO plays a crucial role in the classification of MNIST images. Instead, we now demonstrate that the emergence of these long-range structures originates from the dataset itself, rather than the reservoir, and that long-range correlations are not essential for effective performance in this task.

First, we eliminated all interactions within the reservoir by disconnecting the thermal coupling between adjacent neuristors. Consequently, each thermal neuristor responds solely to its designated input pixel, devoid of any contextual awareness. After fine-tuning the hyperparameters via Bayesian optimization \cite{bergstra2013making} to maximize classification accuracy, we observed the resultant reservoir dynamics, as depicted in Fig.~\ref{fig:MNIST_comparison}(a). The corresponding avalanche size distribution is illustrated in Fig.\ref{fig:MNIST_comparison}(c), and an animation of these dynamics is available in Supplementary Movie 3. In this configuration, the distribution reveals more defined structures: system-wide avalanches represent the white background, while smaller avalanches delineate the fractures within the digits. Given the absence of inter-neuristor interactions, it is evident that these structures are derived {\it solely} from the dataset. When compared to the original distribution involving interactions, it becomes apparent that internal interactions within the reservoir tend to smooth out smaller avalanches while retaining the principal structures inherent to the dataset.

Upon training the output layer, the classification accuracy achieved on the MNIST dataset's test set is presented in Fig.~\ref{fig:MNIST_comparison}(d). Remarkably, even in the absence of interactions within the reservoir, we attained an accuracy of 95.8\%, which is nearly equivalent to the performance under the optimized setup that included interactions. This outcome substantiates the assertion that LRO within the reservoir is not essential for achieving high computational effectiveness.

In another experiment, we aimed to minimize LRO in the neuristor array as much as possible. Previous findings, as discussed in the main text, suggest that LRO is influenced by the separation of time scales. To address this, we attempted to reduce the memory within the system and eliminate the slower time scale by decreasing the resistance in the insulating state of the VO$_2$ device. This adjustment involved increasing the ambient temperature, $T_0$, from 325 K to 330 K. Fig.~\ref{fig:timescale} illustrates the impact of this temperature increase on a single neuristor: the VO$_2$ device begins closer to the metallic state with reduced resistance, and after each spiking event, it resets to a lower resistance state. Further increase of $T_0$ ultimately leads the neuristors to remain in the metallic state, thereby ceasing to spike.

As illustrated in Fig.\ref{fig:timescale}(d), the maximum resistance in the insulating state during spiking, denoted as $R_\mathrm{ins}$, is approximately $7\mathrm{k}\Omega$. This value results in an insulating time constant, $\tau_\mathrm{ins}=R_\mathrm{ins}C\sim 1\mu \mathrm{s}$, significantly reduced from $\tau_\mathrm{ins}=7.57\mu\mathrm{s}$ mentioned in the main text. Consequently, both memory and LRO within the system are diminished. Furthermore, as shown in Fig.\ref{fig:timescale}(c), the spiking frequency increases, while the amplitudes of the spikes decrease. This alteration in dynamics leads to a spiking pattern that is more susceptible to disruption by external perturbations. 

Once again, we optimized the hyperparameters using Bayesian optimization \cite{bergstra2013making} to achieve the best classification accuracy on the MNIST dataset. Fig.\ref{fig:MNIST_comparison}(a) captures snapshots of the reservoir dynamics, with an accompanying animation available in Supplemental Movie 4. These visuals demonstrate how the digit patterns rapidly blur and fade, indicating a rapid loss of memory within the system. Fig.\ref{fig:MNIST_comparison}(b) shows that synchronized system-wide spikes from uniform input confirm the reservoir's operation in a rigid state under these conditions. However, when using the MNIST dataset as input, the avalanche size distribution, depicted in Fig.\ref{fig:MNIST_comparison}(c), shows a power-law distribution for smaller avalanches and system-wide activities, mirroring the original setup. As noted previously, this pattern mostly originates from structural features inherent in the dataset, particularly since the modified reservoir struggles to maintain LRO. The classification accuracy recorded on the test set, as depicted in Fig.\ref{fig:MNIST_comparison}(d), ultimately reaches 94.2\%. Although the classification task is executed effectively, the neuristors’ unstable spiking behavior remains less than ideal for achieving higher accuracy.

From the two experiments described above, it is clear that LRO may not necessarily emerge from the reservoir itself, rather it can  be inherited from the dataset. Across all three experiments, there was no discernible correlation between LRO and computational performance. While the original setup, which included the possibility of LRO, performed the best, the non-interacting setup yielded nearly comparable results without any long-range interactions. Meanwhile, the experiment with reduced memory, although showing the lowest performance, still exhibited similar long-range structures as the original setup. Therefore, we conclude that LRO is not essential for effective computational performance in these scenarios.

\subsection{Predicting chaotic dynamics with reservoir computing}

\begin{figure*}
    \centering
    \includegraphics[width=1\linewidth]{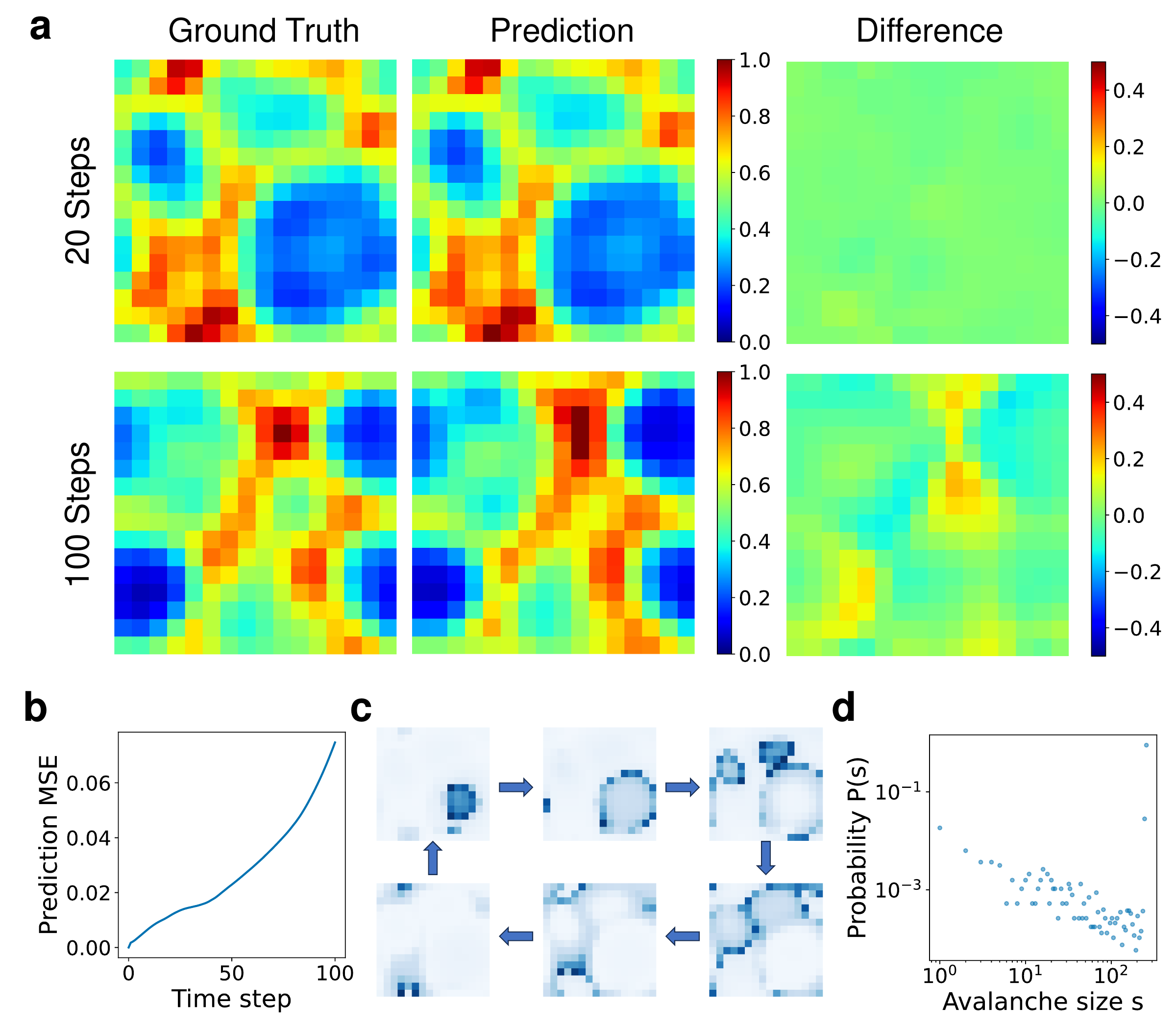}
    \caption{Predicting the chaotic dynamics governed by the 2D Kuramoto-Sivashinsky (KS) equations using reservoir computing. (a) Comparison of the ground truths and predictions, after 20 and 100 time steps, with each time step corresponding to $\Delta t=0.05$ (arbitrary unit). The two rightmost panels illustrate the differences between the predictions and the ground truths, highlighting the model's accuracy over time. (b) The mean-square-error (MSE) of the prediction as a function of the number of time steps predicted. Although the prediction MSE gradually increases with more extended simulation of the dynamics, the predictions remain reasonably accurate for up to 100 time steps, equivalent to $\Delta t=5$ (arbitrary unit). (c) An example of reservoir dynamics is presented, showing patterns that loosely resemble the evolution observed in the KS equation dynamics. (d) Avalanche size distribution with KS dynamics as input to the reservoir. A prominent peak at the right end of the distribution highlights the predominance of system-wide avalanches, confirming that the reservoir operates in a rigid state.
    }
    \label{fig:KS}
\end{figure*}

In this section, we describe an experiment designed to predict chaotic dynamics governed by the 2D Kuramoto-Sivashinsky (KS) equations \cite{kalogirou2015depth} using a reservoir computing framework implemented with a thermal neuristor array.

The 2D KS equation is expressed as:
\begin{equation}
    \frac{\partial u}{\partial t} + \frac{1}{2}|\nabla u|^2 + \Delta u + \Delta^2 u=0 \label{eq:KS}
\end{equation}
where the boundary conditions are spatially periodic. This equation has been extensively studied \cite{kalogirou2015depth} and is known for its chaotic behavior, which poses significant challenges in predicting long-term dynamics.

We discretized Eq.~\eqref{eq:KS} on a $16 \times 16$ square lattice with a unit bond length and simulated the dynamics numerically using a 4th order Runge-Kutta method with a time step of $\Delta t = 0.05$. The simulation began from random initial conditions and, after allowing for the decay of initial transients, the dynamics of the $u$ field were recorded as training data.

To predict the chaotic dynamics described by the KS equation, we utilized reservoir computing with an array of thermal neuristors. This setup is similar to that used in the MNIST classification experiment detailed in the main text, allowing for a direct comparison of the techniques' effectiveness across different types of machine learning tasks.

In our approach, noting that the mean value of the $u$ field is nonzero and decreases over time, yet the KS equation \eqref{eq:KS} depends only on the gradient and not the magnitude of $u$ \cite{kalogirou2015depth}, we performed a preprocessing step. This involved subtracting the mean from the $u$ field and normalizing it between 0 and 1 to yield the transformed field, $\tilde{u}$. Snapshots of the $\tilde{u}$ field are shown in Fig.~\ref{fig:KS}(a). Each $\tilde{u}(x, y)$ value is then linearly transformed into an input voltage for the corresponding thermal neuristor, with the spiking dynamics of the neuristor array serving as the output feature from the reservoir. An output layer is subsequently trained to predict the incremental change $u(x, y, t+\Delta t) - u(x, y, t)$.

Reservoir hyperparameters were optimized using the hyperopt library \cite{bergstra2013making}. The optimized parameters included $C_{\mathrm{th}}=1.073$, noise strength $\sigma=0$, and the transformation formula $V^{\mathrm{in}}=(11.83 - 0.48\tilde{u})$V. Notably, the optimization yielded zero noise, underscoring that noise is detrimental in this reservoir computing task.  %and confirming that placing the neuristor array in the rigid phase enhances performance. 

To enhance performance on this dataset, we implemented several modifications to the reservoir:
\begin{enumerate}
\item The output from the reservoir consists of the magnitudes of the current spikes, rather than merely indicating the presence of a spike.
\item Reflecting the local interaction nature of Eq.~\eqref{eq:KS}, we replaced a fully-connected output layer with a convolution-like layer, where each output neuron is connected to the $5\times 5$ nearest neighbors in the reservoir.
\item We eliminated the softmax nonlinearity in the output layer and employed linear regression for training.
\end{enumerate}

This modified procedure predicts $u(x, y, t+\Delta t)$ from $u(x, y, t)$ at time $t$. Long-time predictions are iteratively performed by using the predicted $u$ field as the input for the subsequent prediction step. The prediction results after 20 and 100 steps with $\Delta t=0.05$ are illustrated in Fig.~\ref{fig:KS}(a), and the mean-square-error of the prediction is plotted in Fig.~\ref{fig:KS}(b), demonstrating good agreement with the ground truth. An example of the dynamics within the reservoir is depicted in Fig.~\ref{fig:KS}(c), which loosely resembles the patterns in the $\tilde{u}$ field.

To verify the presence of LRO, we calculated the avalanche size distribution within the reservoir using the transformed $\tilde{u}$ field from the KS dynamics as input. The results, depicted in Fig.~\ref{fig:KS}(d), are characterized predominantly by system-wide avalanches, indicating that the system maintains a rigid state even with non-uniform inputs. Moreover, the implementation of a locally connected output layer demonstrates that local information alone is sufficient for predicting dynamics, further reinforcing the notion that LRO is not necessary for achieving optimal computational performance.

\end{document}